\documentclass[letterpaper]{article} 
\usepackage{aaai25_arxiv}  
\usepackage{times}  
\usepackage{helvet}  
\usepackage{courier}  
\usepackage[hyphens]{url}  
\usepackage{graphicx} 
\usepackage{natbib}  
\usepackage{caption} 
\usepackage{algorithm}
\usepackage{algorithmic}

\usepackage{newfloat}
\usepackage{listings}
\DeclareCaptionStyle{ruled}{labelfont=normalfont,labelsep=colon,strut=off} 
\floatstyle{ruled}
\newfloat{listing}{tb}{lst}{}
\floatname{listing}{Listing}
\usepackage{hyperref}
\usepackage{amsthm}
\usepackage{algorithm}
\usepackage{algorithmic}
\usepackage{mathtools}
\usepackage{amsfonts}
\newcommand{\bs}{\boldsymbol}
\newcommand{\mc}{\mathcal}

\newcommand{\rmd}{{\rm d}}
\newcommand{\sumi}{\Sigma_i}
\newcommand{\sumj}{\Sigma_j}
\newtheorem{theorem}{Theorem}

\newtheorem{definition}{Definition}
\newtheorem{example}{Example}
\usepackage{comment}
\usepackage{bm}

\title{Synchronization in Learning in Periodic Zero-Sum Games Triggers \\ Divergence from Nash Equilibrium}
\author{
    Yuma Fujimoto\textsuperscript{\rm 1,2,3 $\dagger$},
    Kaito Ariu\textsuperscript{\rm 1},
    Kenshi Abe\textsuperscript{\rm 1,4}
}
\affiliations{
    \textsuperscript{\rm 1}CyberAgent \\
    \textsuperscript{\rm 2}University of Tokyo \\
    \textsuperscript{\rm 3}Soken University \\
    \textsuperscript{\rm 4}University of Electro-Communications \\
    \textsuperscript{$\dagger$}fujimoto.yuma1991@gmail.com
}

\usepackage{bibentry}

\begin{document}

\maketitle

\begin{abstract}
Learning in zero-sum games studies a situation where multiple agents competitively learn their strategy. In such multi-agent learning, we often see that the strategies cycle around their optimum, i.e., Nash equilibrium. When a game periodically varies (called a ``periodic'' game), however, the Nash equilibrium moves generically. How learning dynamics behave in such periodic games is of interest but still unclear. Interestingly, we discover that the behavior is highly dependent on the relationship between the two speeds at which the game changes and at which players learn. We observe that when these two speeds synchronize, the learning dynamics diverge, and their time-average does not converge. Otherwise, the learning dynamics draw complicated cycles, but their time-average converges. Under some assumptions introduced for the dynamical systems analysis, we prove that this behavior occurs. Furthermore, our experiments observe this behavior even if these assumptions are removed. This study discovers a novel phenomenon, i.e., synchronization, and gains insight widely applicable to learning in periodic games.
\end{abstract}

%
\begin{links}
    \link{Code}{https://github.com/CyberAgentAILab/periodic_games_synchronization}
\end{links}

\section{Introduction}
Learning in games discusses how multiple agents optimize their strategies in the repetition of games~\cite{fudenberg1998theory}. Their optimal strategies are usually characterized by Nash equilibrium~\cite{nash1950equilibrium}, where every player cannot increase its payoff by other strategies. When their payoffs compete with each other, i.e., in zero-sum games, it is difficult to achieve equilibrium by naive learning algorithms, such as gradient descent-ascent (GDA). This is because one's optimal strategy crucially depends on the other's strategy. Their strategies typically draw a complex trajectory, such as a cycle around the equilibrium, without convergence. Thus, dynamical systems analysis is often introduced in learning in games to understand such complex behavior~\cite{sato2002chaos, piliouras2014persistent, bloembergen2015evolutionary, mertikopoulos2016learning, mertikopoulos2018cycles, bailey2019multi, fujimoto2024global, fujimoto2024nash}.

Learning in games typically assumes that the same game is repeatedly played, and the dynamical systems in learning (say, learning dynamics) may be even more complex when the game changes with time (called ``time-varying'' games). These time-varying games currently attract much attention~\cite{fiez2021online, zhang2022no, duvocelle2023multiagent, yan2023fast, anagnostides2023convergence, feng2023last, feng2024last}. One of the major factors causing such time-varyingness is the periodic effect in the real-world environment, such as daily and seasonal cycles. The games that change periodically are called ``periodic'' games. Although such periodic games are significant, they have been studied only in a few papers~\cite{fiez2021online, feng2023last, feng2024last}. Moreover, these studies focus on a special class of periodic games whose equilibrium is invariant over time. Since the Nash equilibrium is the target of learning, the movement of the equilibrium is expected to impact the learning dynamics significantly\footnote{Indeed, it was said that ``An interesting question arises --- in periodic games where there is no common equilibrium.'' in~\cite{feng2024last}}. To summarize, the learning dynamics in general periodic games with their Nash equilibrium time-varying are crucial but still unexplored.

Besides the trajectory of the learning dynamics, the behavior of the ``time-average'', defined as the average of the trajectory, is also of interest. In time-invariant zero-sum games, this time-average is known to converge to the Nash equilibrium over time. This convergence property is intuitive because, in GDA, the trajectory cycles around the Nash equilibrium, and the deviations from the equilibrium cancel each other in the long run. However, since the learning dynamics are expected to be complex in periodic games equipped with time-varying equilibrium, this time-average convergence becomes a non-trivial problem. The dynamical systems analysis is also helpful to examine the time-average convergence, which has been usually proved by the no-regret property~\cite{banerjee2005efficient, zinkevich2007regret, daskalakis2011near}.

We tackle a dynamical systems analysis in general periodic games and discuss the time-average convergence. To the best of our knowledge, this is the first study to focus on the phenomenon triggered by the synchronization of learning with periodic games. Interestingly, when the speeds of learning and time-varying games synchronize, the players' strategies diverge from the Nash equilibrium, and their time-average does not converge. Otherwise, they cycle with complicated trajectories, and their time-average converges. We practically show that this phenomenon is intrinsic by giving a simple example, i.e., periodic matching pennies. We prove that this phenomenon holds in $2\times 2$ matrix games with arbitrary wave shapes of periodic games. Furthermore, our experiments support that this phenomenon is valid in more general settings: 1) independent of the action numbers, 2) with the boundary constraint of strategy spaces. This study finds a novel phenomenon in learning in periodic games and extracts insight to understand the learning dynamics theoretically.

\section{Setting}
\subsection{Normal-Form Games}
Let us introduce normal-form games between two players, ${\sf X}$ and ${\sf Y}$. Every round, they independently choose their actions from given sets, $\mc{A}=\{a_i\}_{1\le i\le m_{\sf X}}$ and $\mc{B}=\{b_j\}_{1\le j\le m_{\sf Y}}$, respectively. When they choose $a_i$ and $b_j$, they receive the payoffs of $u_{ij}$ and $v_{ij}$, respectively. We define their payoff matrices as $\bs{U}:=(u_{ij})_{ij}$ and $\bs{V}:=(v_{ij})_{ij}$. We assume that the summation of their payoffs is zero, i.e., $\bs{V}=-\bs{U}$.

In normal-form games, players choose their actions following their strategies, denoted as $\bs{x}=(x_{i})_{i}\in\mc{X}:=\Delta^{m_{\sf X}-1}$ and $\bs{y}=(y_{j})_{j}\in\mc{Y}:=\Delta^{m_{\sf Y}-1}$. Here, $x_{i}$ means the probability that ${\sf X}$ chooses action $a_i$. ${\sf X}$'s expected payoff is given by $u(\bs{x},\bs{y}):=\bs{x}^{\rm T}\bs{U}\bs{y}=\sumi\sumj x_iy_ju_{ij}$. By the definition of zero-sum games, ${\sf Y}$'s expected payoff is computed as $v(\bs{x},\bs{y})=-u(\bs{x},\bs{y})$.

\subsection{Periodic Games}
This study considers a situation where the game can fluctuate periodically. Thus, the payoff matrix depends on time $\bs{U}=\bs{U}(t)$ under the following constraint.

\begin{definition}[Periodic game]
A periodic game satisfies $\bs{U}(t)=\bs{U}(t+2\pi/\omega)$ with some $\omega\in\mathbb{R}$ for all $t$.
\end{definition}

Here, the period is $2\pi/\omega$, where $\omega$ represents the frequency of the periodic game.

\subsection{Gradient Descent-Ascent}
Learning in games discusses a process where each player sequentially optimizes their strategies with time $t$. One of the representative learning algorithms is GDA, described as
\begin{align}
    \bs{x}&=\arg\max_{\bs{x}\in\mc{X}}\left[\bs{x}\cdot\bs{x}^{\dagger}-\frac{\|\bs{x}\|_{2}^2}{2}\right], &\dot{\bs{x}}^{\dagger}&=+\frac{\partial u(\bs{x},\bs{y})}{\partial\bs{x}},
    \label{x_FTRL_EU} \\
    \bs{y}&=\arg\max_{\bs{y}\in\mc{Y}}\left[\bs{y}\cdot\bs{y}^{\dagger}-\frac{\|\bs{y}\|_{2}^2}{2}\right], &\dot{\bs{y}}^{\dagger}&=-\frac{\partial u(\bs{x},\bs{y})}{\partial\bs{y}}.
    \label{y_FTRL_EU}
\end{align}
This algorithm is also known as the Follow the Regularized Leader (FTRL) with the Euclidean regularizer.

In the interior of the strategy space, Eqs.~\eqref{x_FTRL_EU} and~\eqref{y_FTRL_EU} are rewritten as
\begin{align}
    &\frac{\rmd}{\rmd t}\begin{pmatrix}
        \bs{x} \\
        \bs{y} \\
    \end{pmatrix}=\begin{pmatrix}
        \bs{O}_{\sf X} & +\bs{P}_{\sf X}\bs{U} \\
        -\bs{P}_{\sf Y}\bs{U}^{\rm T} & \bs{O}_{\sf Y} \\
    \end{pmatrix}\begin{pmatrix}
        \bs{x} \\
        \bs{y} \\
    \end{pmatrix},
    \label{cFTRL_EU}\\
    &\bs{P}_{\sf X}:=\bs{I}_{\sf X}-\frac{\bs{1}_{\sf X}\bs{1}_{\sf X}^{\rm T}}{m_{\sf X}}, \quad \bs{P}_{\sf Y}:=\bs{I}_{\sf Y}-\frac{\bs{1}_{\sf Y}\bs{1}_{\sf Y}^{\rm T}}{m_{\sf Y}},
\end{align}
(see Appendix~B.1 for derivation). Here, the subscripts of ${\sf X}$ and ${\sf Y}$ mean that the dimension of the vector or matrix is $m_{\sf X}$ and $m_{\sf Y}$, respectively. Furthermore, $\bs{O}_{\sf X}$ and $\bs{O}_{\sf Y}$ are zero matrices, $\bs{I}_{\sf X}$ and $\bs{I}_{\sf Y}$ are identity matrices, and $\bs{1}_{\sf X}$ and $\bs{1}_{\sf Y}$ are all-ones vectors. Here, $\bs{P}_{\sf X}$ and $\bs{P}_{\sf Y}$ mean the projection to the strategy space by the Euclidean regularizer.

\subsection{Eigenvalue of Learning}
Let us analyze Eq.~\eqref{cFTRL_EU}. For simplicity, we assume that the payoff matrix $\bs{U}$ is time-invariant. Since Eq.~\eqref{cFTRL_EU} is a linear dynamical system~\cite{hirsch2013differential}, its solution is characterized by the eigenvalues of
\begin{align}
    \bs{A}:=\begin{pmatrix}
        \bs{O}_{\sf X} & +\bs{P}_{\sf X}\bs{U} \\
        -\bs{P}_{\sf Y}\bs{U}^{\rm T} & \bs{O}_{\sf Y} \\
    \end{pmatrix}.
    \label{matrix_A}
\end{align}
Here, note that if the projection operator ($\bs{P}_{\sf X}$ and $\bs{P}_{\sf Y}$) is removed, $\bs{A}$ is a skew-Hermitian matrix, all of which eigenvalues are known to zero or purely imaginary. Interestingly, this projection operator does not affect the property of eigenvalues, and we can prove that all the eigenvalues of $\bs{A}$ are zero or purely imaginary (see Appendix~A for detailed calculation).

For the convenience of notation, consider $m_{\sf X}=m_{\sf Y}=:m$, then the eigenvalues of $\bs{A}$ are denoted as $\{\pm i\alpha^{(j)}|j=1,2,\cdots,m\}$ with $\alpha^{(j)}\in\mathbb{R}$ (purely imaginary eigenvalues). We can rearrange them so that they satisfy $\alpha^{(1)}\ge\alpha^{(2)}\ge\cdots\ge\alpha^{(m)}=0$. Here, at least one pair of the eigenvalues is zero ($\alpha^{(m)}=0$) because both $\bs{P}_{\sf X}\bs{U}$ and $\bs{P}_{\sf Y}\bs{U}^{\rm T}$ degenerate in $\bs{A}$. Now, these purely imaginary eigenvalues mean that the dynamics of Eq.~\eqref{cFTRL_EU} are given by the summation of circular functions. In particular, $\alpha^{(j)}$ corresponds to the cycling speed of each circular function. We simply call $\alpha^{(j)}$ as ``eigenvalue'' throughout this paper. We remark that the cycling behavior characterized by purely imaginary eigenvalues is consistent with the time-invariance of distance from the Nash equilibrium~\cite{piliouras2014persistent} and the Hamiltonian structure~\cite{bailey2019multi}.

\section{Example: Eigenvalue Invariant Game}
In order to intuitively understand the behavior of learning in periodic games, we now consider the simplest periodic game, named eigenvalue invariant game, which provides a phenomenon beyond an existing class of games equipped with the time-invariant Nash equilibrium~\cite{fiez2021online, feng2023last, feng2024last}. This eigenvalue invariant game is given by two $2\times 2$ matrices of $\bar{\bs{U}}\in\mathbb{R}^{2\times 2}$ and $\Delta\bs{U}\in\mathbb{R}^{2\times 2}$ and formulated as follows.

\begin{example}[Eigenvalue invariant game]
\label{Exm_simplest}
For given two payoff matrices $\bar{\bs{U}}=(\bar{u}_{ij})_{ij}\in\mathbb{R}^{2\times 2}$ and $\Delta\bs{U}=(\Delta u_{ij})_{ij}\in\mathbb{R}^{2\times 2}$ with the constraint of $\Delta u_{11}-\Delta u_{12}-\Delta u_{21}+\Delta u_{22}=0$, the eigenvalue invariant game $\bs{U}(t)$ is formulated as
\begin{align}
    \bs{U}(t)=\bar{\bs{U}}+\Delta\bs{U}\cos\omega t.
\end{align}
\end{example}

Here, $\bar{\bs{U}}$ is the time average of periodic game $\bs{U}(t)$, while $\Delta\bs{U}$ is the amplitude of the oscillation. Thus, this periodic game oscillates between $\bar{\bs{U}}+\Delta\bs{U}$ and $\bar{\bs{U}}-\Delta\bs{U}$. The constraint for $\Delta\bs{U}$ means that the eigenvalue is time-invariant, as shown later.

\subsection{Analysis of Eigenvalue Invariant Game} \label{Subsec_analysis_eigenvalue}
We now analyze learning in these eigenvalue invariant games. First, since $\bs{x}=(x_1,x_2)\in\Delta^{1}$ should be satisfied in $2\times 2$ matrix games, $\bs{x}$ can be described by a single value $x\in[0,1]$ as $\bs{x}=(x,1-x)$. In a similar manner, we only consider $y\in[0,1]$ for $\bs{y}=(y,1-y)$. Thus, the learning dynamics is described only by the two variables, $x(t)$ and $y(t)$.

By directly calculating the eigenvalues of Eq.~\eqref{matrix_A}, the eigenvalues of learning are obtained as
\begin{align}
    \alpha^{(1)}&=\frac{1}{2}(\bar{u}_{11}-\bar{u}_{12}-\bar{u}_{21}+\bar{u}_{22})=:\alpha,
    \label{ev1}\\
    \alpha^{(2)}&=0,
    \label{ev2}
\end{align}
(see Appendix~B.2 for derivation). Furthermore, the Nash equilibrium of the time-average of the periodic game, denoted as $(\bar{x}^{*},\bar{y}^{*})$, is given by
\begin{align}
    \bar{x}^{*}:=\frac{-\bar{u}_{21}+\bar{u}_{22}}{2\alpha},\quad \bar{y}^{*}:=\frac{-\bar{u}_{12}+\bar{u}_{22}}{2\alpha}.
\end{align}
Next, the deviation of the Nash equilibrium caused by the oscillation of the game, denoted as $(\Delta x^{*},\Delta y^{*})$, is also given by
\begin{align}
    \Delta x^{*}:=\frac{-\Delta u_{21}+\Delta u_{22}}{2\alpha},\quad \Delta y^{*}:=\frac{-\Delta u_{12}+\Delta u_{22}}{2\alpha}.
\end{align}
Thus, the Nash equilibrium oscillates between $(\bar{x}^{*}+\Delta x^{*}, \bar{y}^{*}+\Delta y^{*})$ and $(\bar{x}^{*}-\Delta x^{*}, \bar{y}^{*}-\Delta y^{*})$, which is beyond the scope of the previous studies~\cite{fiez2021online, feng2023last, feng2024last}.

Finally, the learning dynamics, i.e., Eq.~\eqref{cFTRL_EU}, are described as
\begin{align}
    \dot{x}(t)&=+\alpha(y(t)-\bar{y}^*-\Delta y^*\cos\omega t),
    \label{x_dyn_simplest}\\
    \dot{y}(t)&=-\alpha(x(t)-\bar{x}^*-\Delta x^*\cos\omega t).
    \label{y_dyn_simplest}
\end{align}
Interestingly, these equations correspond to the dynamics of forced pendulum without dissipation~\cite{mawhin2004global}. In other words, the competitive learning between the players can be interpreted as an oscillation inherent in a pendulum, whereas the vibration of the Nash equilibrium can be as the external force into the pendulum.

\begin{figure*}[h!]
    \centering
    \includegraphics[width=0.8\hsize]{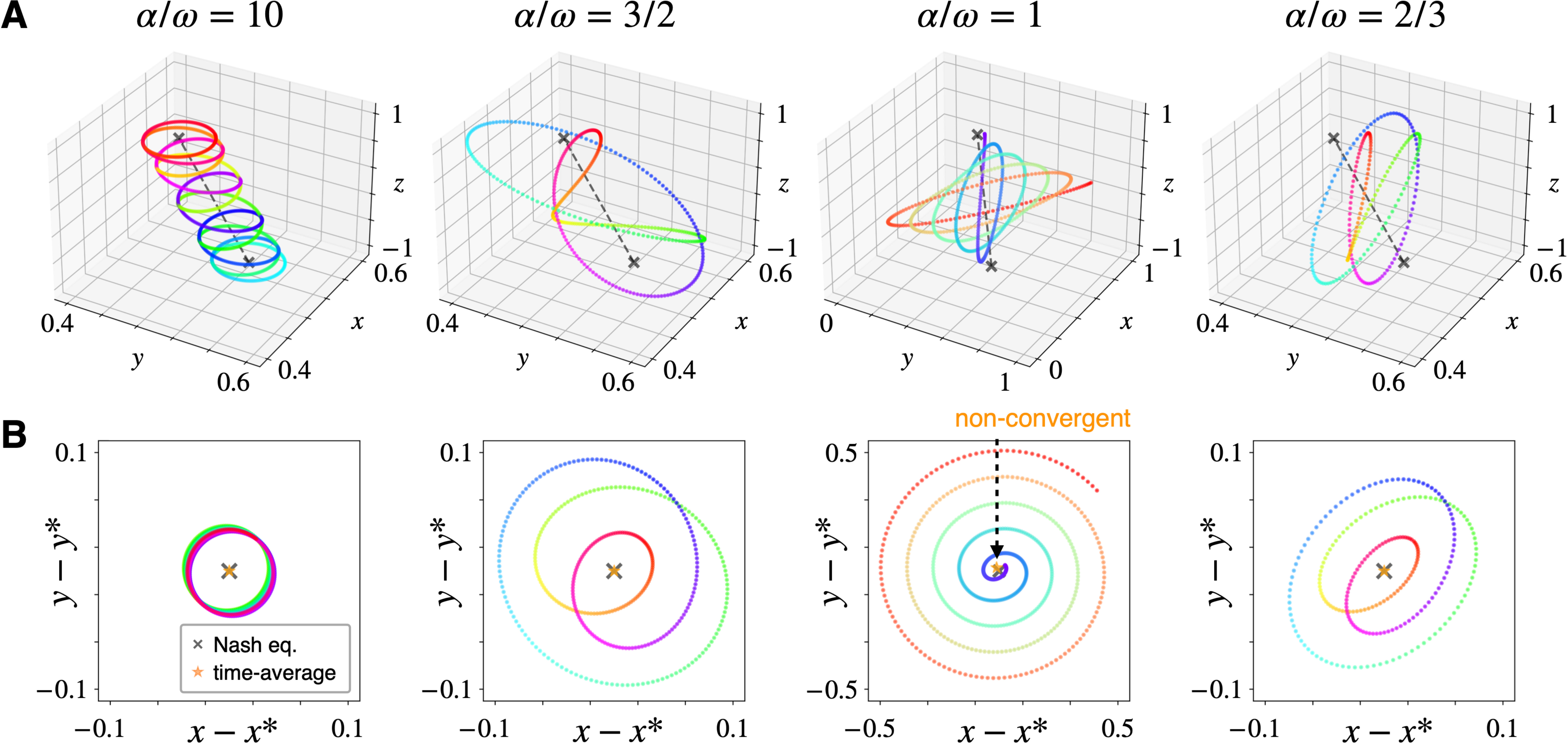}
    \caption{Learning dynamics in the periodic game of matching pennies with the time-varying Nash equilibrium. (A) The trajectories of the learning dynamics. The panels show the cases of $\alpha/\omega=10$, $2/3$, $1$, and $2/3$, from left to right. In each panel, the x-, y-, z-axes indicate $x(t)$, $y(t)$, and $z(t)$, respectively. The black broken line shows the trajectory of the Nash equilibrium, i.e., $(x^{*}(t),y^{*}(t))$, and the cross markers show the edges of the oscillation of this equilibrium. In the cases of $\alpha/\omega\neq 1$, the rainbow color shows the time of a single cycle. In $\alpha/\omega=1$, the rainbow color shows the passing of time from blue to red. (B) The projection of the trajectories of Panels A to the plane of $x(t)-x^{*}(t)$ and $y(t)-y^{*}(t)$. In each panel, the black cross marker shows the projection of the Nash equilibrium. The color corresponds to that of Panels A. The orange star markers show the time-average value of the plotted trajectory. The value does not correspond to the equilibrium in $\alpha/\omega=1$. Otherwise, it corresponds. To simulate the learning dynamics, we use the Runge-Kutta fourth-order method with a step size of $1/40$. The initial strategies are set to the Nash equilibrium of the time-average game.}
    \label{F01}
\end{figure*}

\subsection{Solution of Learning Dynamics}
The learning dynamics, Eqs.~\eqref{x_dyn_simplest} and~\eqref{y_dyn_simplest}, are solvable. In the case of $\alpha/\omega=1$, the solution is
\begin{align}
    x(t)&=\bar{x}^{*}+\frac{1}{2}\Delta x^{*}(\omega t\sin\omega t+\cos\omega t)
    \nonumber\\
    &\hspace{0.3cm}-\frac{1}{2}\Delta y^{*}\omega t\cos\omega t+c_1\cos\alpha t+c_2\sin\alpha t,
    \label{x_sol_simplest1}\\
    y(t)&=\bar{y}^{*}+\frac{1}{2}\Delta y^{*}(\omega t\sin\omega t+\cos\omega t)
    \nonumber\\
    &\hspace{0.3cm}+\frac{1}{2}\Delta x^{*}\omega t\cos\omega t+c_2\cos\alpha t-c_1\sin\alpha t,
    \label{y_sol_simplest1}
\end{align}
where $c_1$ and $c_2$ are uniquely determined by the initial state condition (see Appendix~B.3 for the detailed expression of $c_1$ and $c_2$). This solution means that $x(t)$ and $y(t)$ diverges from the average Nash equilibrium $(\bar{x}^{*},\bar{y}^{*})$ at a speed of $\sqrt{|\Delta x^{*}|^2+|\Delta y^{*}|^2}\omega/2$.

On the other hand, in the case of $\alpha/\omega\neq 1$, the solution is
\begin{align}
    x(t)&=\bar{x}^{*}+\frac{\alpha^2}{\alpha^2-\omega^2}\Delta x^{*}\cos\omega t+\frac{\alpha\omega}{\alpha^2-\omega^2}\Delta y^{*}\sin\omega t
    \nonumber\\
    &\hspace{0.3cm}+c_{1}\cos\alpha t+c_{2}\sin\alpha t,
    \label{x_sol_simplest2}\\
    y(t)&=\bar{y}^{*}+\frac{\alpha^2}{\alpha^2-\omega^2}\Delta y^{*}\cos\omega t-\frac{\alpha\omega}{\alpha^2-\omega^2}\Delta x^{*}\sin\omega t
    \nonumber\\
    &\hspace{0.3cm}+c_{2}\cos\alpha t-c_{1}\sin\alpha t.
    \label{y_sol_simplest2}
\end{align}
Here, $x(t)$ and $y(t)$ consist of two periodic solutions with the frequencies of $\alpha$ and $\omega$. Thus, the solution qualitatively changes depending on whether $\alpha/\omega=1$ or $\neq 1$.

\subsection{Time-Average Analysis}
Under fixed zero-sum payoff matrices, the dynamics of GDA are known to converge to the Nash equilibrium (called time-average convergence). We now discuss time-average convergence under periodic payoff matrices. The time-average of X's strategy until time $T>0$ is defined as
\begin{align}
    \bar{x}(T):=\frac{1}{T}\int_{0}^{T}x(t)\rmd t.
\end{align}
Interestingly, depending on whether $\alpha/\omega=1$ or $\neq 1$, the time-average has different properties, as shown in the following. See Appendix~C for the full proofs of all the theorems.

\begin{theorem}[Time-average cycling in $\alpha/\omega=1$]
\label{Thm_div_simplest}
In the eigenvalue invariant games, when $\alpha/\omega=1$ holds, $\bar{x}(T)$ and $\bar{y}(T)$ cycle around the Nash equilibrium.
\end{theorem}

{\it Proof sketch.} This theorem is proved by directly calculating the integral of Eqs.~\eqref{x_sol_simplest1} and~\eqref{y_sol_simplest1}. Intuitively, since the solution diverges as it oscillates, its time-average is influenced by $x(t)$ of large time $t\simeq T$ and never converges.
\qed

\begin{theorem}[Time-average convergence in $\alpha/\omega\neq 1$]
\label{Thm_conv_simplest}
In the eigenvalue invariant games, when $\alpha/\omega\neq 1$ holds, $\bar{x}(T)$ and $\bar{y}(T)$ converge to the Nash equilibrium.
\end{theorem}

{\it Proof sketch.} This theorem is proved by directly calculating the integral of Eqs.~\eqref{x_sol_simplest2} and~\eqref{y_sol_simplest2}, where each cycle is canceled out.
\qed

\subsection{Visualization and Interpretation}
Let us visualize the learning dynamics (see Fig.~\ref{F01}) and interpret their behavior. This section focuses on a periodic game formulated by $\bar{\bs{U}}=((1,-1),(-1,1))$ and $\Delta\bs{U}=((1/10,0),(0,-1/10))$ in Exm.~\ref{Exm_simplest}. Here, $\bar{\bs{U}}$ shows Matching Pennies with the only equilibrium of $(\bar{x}^{*},\bar{y}^{*})=(1/2,1/2)$. This periodic game has a time-invariant eigenvalue of $\alpha=2$. However, its Nash equilibrium oscillates with the amplitude of $(\Delta x^{*},\Delta y^{*})=(-1/40,-1/40)$.

First, let us see the case of $\alpha/\omega=10$. This case roughly means that the players learn ten times faster than the game changes. In other words, it approximately corresponds to a classical situation where a game is time-invariant. Thus, the learning dynamics are also approximated as the classical dynamics, where the strategies of both players almost cycle around the Nash equilibrium.

Next, we see the case of $\alpha/\omega=3/2$, where the speed of game change is comparable but not equal to that of learning. In this case, the learning interacts with the game change and provides complex dynamics (see Panel A). Although the trajectory cycles around the Nash equilibrium many times, the distance from the Nash equilibrium gets closer and farther as the equilibrium moves. After a while, the trajectory returns to its initial state and thus is periodic as described by Eqs.~\eqref{x_sol_simplest2} and~\eqref{y_sol_simplest2}. In addition, the average of the trajectory corresponds to the equilibrium (see Panel B). The above applies also to the case when the game changes faster than the players learn (e.g., $\alpha/\omega=2/3$ in the panels).

Finally, let us focus on the case of $\alpha/\omega=1$, where the speed of game change is equal to that of learning. In other words, the cycling behavior by learning synchronizes the oscillation of the game. Thus, each time the game oscillates once, the amplitude of the cycle is amplified a little (see Panel A). By this amplification accumulated, the trajectory diverges from the Nash equilibrium as expressed in Eqs.~\eqref{x_sol_simplest1} and~\eqref{y_sol_simplest1}. Furthermore, the average of the trajectory does not correspond to the Nash equilibrium.


\begin{figure*}[h!]
    \centering
    \includegraphics[width=0.65\hsize]{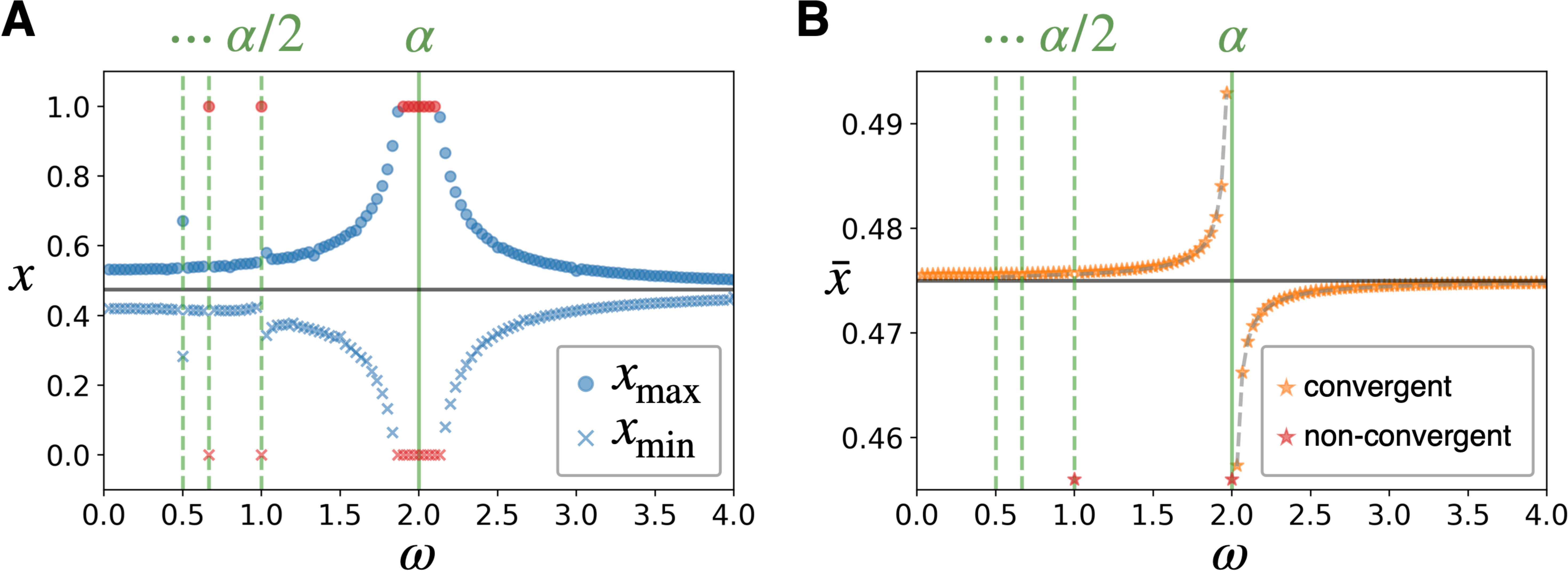}
    \caption{Experiments for learning dynamics in the eigenvalue invariant game for $\bar{\bs{U}}=((1.1,-1),(-1,0.9))$, $\Delta\bs{U}=((0.2,0),(0,0))$ without the constraint of their eigenvalues fixed. (A) The maximum (circle) and minimum (cross) values of $x(t)$ within $0\le t\le T=3\times 10^{4}$. Here, the blue markers indicate that the value exists in the interior of $[0,1]$, while the red exterior. (B) The time-average $\bar{x}$ in the last time $T$. The orange marker is plotted when the time-average sufficiently converges, i.e., the time-average moves less than $10^{-3}$ in the last $10^{2}$ time. The red marker means that the time-average does not converge and take the value outside of the y-axis. The gray broken line shows the analytical solution. The numerical method and parameter is the same as Fig.~\ref{F01}.}
    \label{F02}
\end{figure*}

\section{Theory on Eigenvalue Varying Games}
We now extend the insight obtained from the eigenvalue invariant game (i.e., Exm.~\ref{Exm_simplest}) to more general settings. The eigenvalue invariant game is generalized in the following two points: 1) its eigenvalue time-varying and 2) the wave shape. In the definition, we assume the smoothness of the wave shape to take the time integral of the payoff matrix.

\begin{definition}[$2\times 2$ smooth periodic games] \label{Def_2x2smooth}
$2\times 2$ smooth periodic games are defined as any smooth $\{\bs{U}(t)\}_{0\le t<2\pi/\omega}\in \mathbb{R}^{2\times 2}\times [0,2\pi/\omega)$, satisfying $\bs{U}(t)=\bs{U}(t+2\pi/\omega)$.
\end{definition}

\subsection{Solution of Learning Dynamics}
Below, we use the notation of
\begin{align}
    2f(t)&:=u_{11}(t)-u_{12}(t)-u_{21}(t)+u_{22}(t), \\
    2g(t)&:=-u_{12}(t)+u_{22}(t), \\
    2h(t)&:=-u_{21}(t)+u_{22}(t).
\end{align}
All of these functions are periodic with frequency $\omega$. Here, we see that $f(t)$ represents the eigenvalue of learning and varies over time. By using these equations, we describe the learning dynamics in Def.~\ref{Def_2x2smooth} as
\begin{align}
    \dot{x}(t)&=+f(t)y(t)-g(t),
    \label{dx_general}\\
    \dot{y}(t)&=-f(t)x(t)+h(t).
    \label{dy_general}
\end{align}
These simultaneous differential equations are solved as
\begin{align}
    x(t)&=\int_{0}^{t}h(\tau)\sin(F(t)-F(\tau))\rmd\tau
    \nonumber\\
    &\hspace{0.3cm}-\int_{0}^{t}g(\tau)\cos(F(t)-F(\tau))\rmd\tau
    \nonumber\\
    &\hspace{0.3cm}+c_1\cos F(t)+c_2\sin F(t),
    \label{x_sol_general}\\
    y(t)&=\int_{0}^{t}g(\tau)\sin(F(t)-F(\tau))\rmd\tau
    \nonumber\\
    &\hspace{0.3cm}+\int_{0}^{t}h(\tau)\cos(F(t)-F(\tau))\rmd\tau
    \nonumber\\
    &\hspace{0.3cm}+c_2\cos F(t)-c_1\sin F(t).
    \label{y_sol_general}
\end{align}
Here, $c_{1}$ and $c_{2}$ are uniquely determined by the initial state condition (see Appendix~B.4 for detailed expression of $c_1$ and $c_2$). In addition, we defined $F(t):=\int_{0}^{t}f(\tau)\rmd\tau$. This $F(t)$ is not periodic, but other than the linear term $\Delta F(t):=F(t)-\bar{f}t$ is periodic with frequency $\omega$. This linear term corresponds to the average eigenvalue, i.e., $\alpha=\bar{f}$.

\subsection{Time-Average Analysis}
Let us analyze the time-average of the learning dynamics, i.e., Eqs.~\eqref{x_sol_general} and~\eqref{y_sol_general}. Since we assume general smooth periodic functions for $f$, $g$, and $h$, we cannot analytically calculate the integral of $x(t)$, as different from the case of Exm.~\ref{Exm_simplest}. Thus, we have to devise a way to evaluate $x(T)$ (see the following proof sketch for details). Now, when $\alpha/\omega\in\mathbb{N}$ holds, we prove that the time-average diverges generically (see Appendix~D for some exceptions).

\begin{theorem}[Time-average divergence in $\alpha/\omega\in\mathbb{N}$]
\label{Thm_div_general}
In $2\times 2$ smooth periodic games with $\alpha/\omega\in\mathbb{N}$, $\bar{x}(T)$ and $\bar{y}(T)$ diverge over time $T$.
\end{theorem}

{\it Proof sketch.} As different from the proofs of Thms.~\ref{Thm_div_simplest} and~\ref{Thm_conv_simplest}, the integral of the trajectory (Eqs.~\eqref{x_sol_general} and~\eqref{y_sol_general}) cannot be calculated directly. In the case of $\alpha/\omega\in\mathbb{N}$, however, we can easily evaluate the time-average by paying attention to the periodicity of $\sin F(\cdot)$ and $\cos F(\cdot)$ with frequency $\omega$. We divide the ranges of integrals, i.e., $t\in[0,T]$ and $\tau\in[0,t]$, into the intervals of $2\pi/\omega$, and the time-average is also divided into the following four terms: 1) the term of diverges which is $O(T)$ in the time-average, 2) the term oscillates which is $O(1)$, 3) the term converges which is $O(1)$, and 4) the term is negligible which is $O(1/T)$. In general functions, $f$, $g$, and $h$, the divergence term takes a finite value, and thus we prove that the time-average is $O(T)$ and thus diverges with time.
\qed

Otherwise, the time-average converges as follows.

\begin{theorem}[Time-average convergence in $\alpha/\omega\notin\mathbb{N}$]
\label{Thm_conv_general}
In $2\times 2$ smooth periodic games with $\alpha/\omega\notin\mathbb{N}$, $\bar{x}(T)$ and $\bar{y}(T)$ converge over time $T$.
\end{theorem}

{\it Proof sketch.} We consider the same division of the ranges of the integrals as proof of Thm.~\ref{Thm_div_general} and evaluate the four terms. In the case of $\alpha/\omega\notin\mathbb{N}$, however, this evaluation is much more difficult because we cannot generally use the periodicity of $\sin F(\cdot)$ and $\cos F(\cdot)$. By utilizing a property of the non-periodicity, the divergence term falls into $O(1)$, while the oscillation term into $O(1/T)$. Thus, the time-average converges with neither divergence nor oscillation.
\qed

\begin{figure*}[h!]
    \centering
    \includegraphics[width=0.85\hsize]{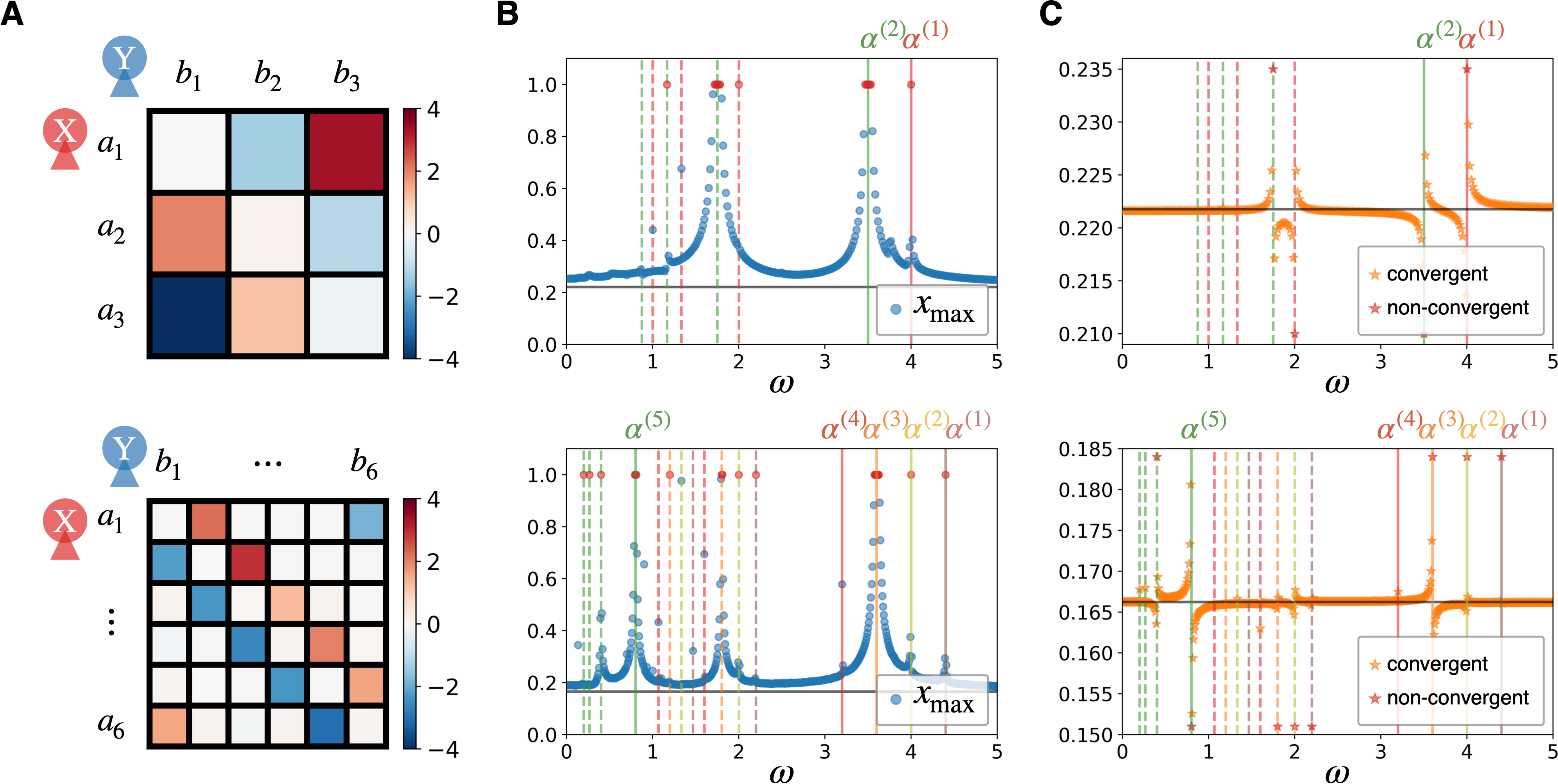}
    \caption{Learning dynamics in games with more than two actions. The upper panels show an example of $3\times 3$ matrix games, while the lower $6\times 6$. (A) $\bar{U}$, i.e., the time-average payoff matrix is visualized. The row ($a_i$) and column ($b_i$) indicate X's and Y's actions, respectively. The darker red shows that X receives more payoff, while the darker blue shows that Y receives more. Here, note that a non-dominant structure is given in each payoff matrix, where each action wins one of the opponent's actions but loses another. We also remark that all the elements of the payoff matrix are non-zero (colored by light red or blue). We further consider the perturbation in the payoff matrix. Each element are independently perturbed by the four waves of $\cos\omega t$, $\sin\omega t$, $\cos2\omega t$, and $\sin2\omega t$ with the amplitude of a random number $\sim\mc{N}(0,0.04^2)$ in the upper panels and $\sim\mc{N}(0,0.02^2)$ in the lower. (B) The maximum value of $x_1(t)$ in overall time $0\le t\le T=3\times 10^{4}$. The meanings of the markers and axes are the same as Fig.~\ref{F02}-A. (C) The average value in the last time $T$. We regard the time-average as convergent when it moves less than $10^{-3}$ in the last $10^{3}$ time. The meanings of the markers and axes are the same as Fig.~\ref{F02}-B. The numerical method and parameter is the same as Fig.~\ref{F01}.}
    \label{F03}
\end{figure*}

\subsection{Experimental Verification for Theory}
Let us see that Thms.~\ref{Thm_div_general} and~\ref{Thm_conv_general} hold in experiments. It is enough to simply extend Exm.~\ref{Exm_simplest} not to satisfy the constraint for $\Delta\bs{U}$, and we use the example of $\bar{\bs{U}}=((1.1,-1),(-1,0.9))$ and $\Delta\bs{U}=((0.2,0),(0,0))$. This example is calculated as $f(t)=2+0.1\cos\omega t$ and $g(t)=h(t)=0.95$, showing that not only the Nash equilibrium but the eigenvalue oscillates. Fig.~\ref{F02} shows comprehensive experiments for various $\omega$.

First, Panel A shows the maximum and minimum values of $x(t)$ for the whole $t$. We see that $x(t)$ always continues to oscillate. Especially in $\alpha/\omega\in\mathbb{N}$, the amplitude of this oscillation is large. Elsewhere, the amplitude is small.

Next, Panel B shows the time-average $\bar{x}(T)$ for sufficiently large $T$. We observe that the time-average diverges in $\alpha/\omega\in\mathbb{N}$, reflecting Thm.~\ref{Thm_div_general}. Here, however, note that when $\alpha/\omega$ takes a sufficiently large integer ($\alpha/\omega\ge 3$ in the panel), the divergence is weak and judged to be convergent. The time-average converges in $\alpha/\omega\notin\mathbb{N}$, reflecting Thm.~\ref{Thm_conv_general}.


\section{Experimental Results}
So far, for theoretical analyses, we have assumed 1) two-action games and 2) no boundary in the strategy spaces. In this section, we experimentally demonstrate that independent of these assumptions, the following ``synchronization phenomenon'' occurs.
\begin{itemize}
\item When the frequency ($\omega$) synchronizes with the eigenvalues $\alpha^{(j)}$, the dynamics diverge from the Nash equilibrium, and their time-average never converges.
\item Otherwise, the dynamics do not diverge, and their time-average converges.
\end{itemize}
As follows, we show experiments 1) for various action numbers and 2) with the boundary constraint of the strategy space. See also Appendix~E.1 for non-smooth periodic games.

\begin{figure*}[h!]
    \centering
    \includegraphics[width=0.8\hsize]{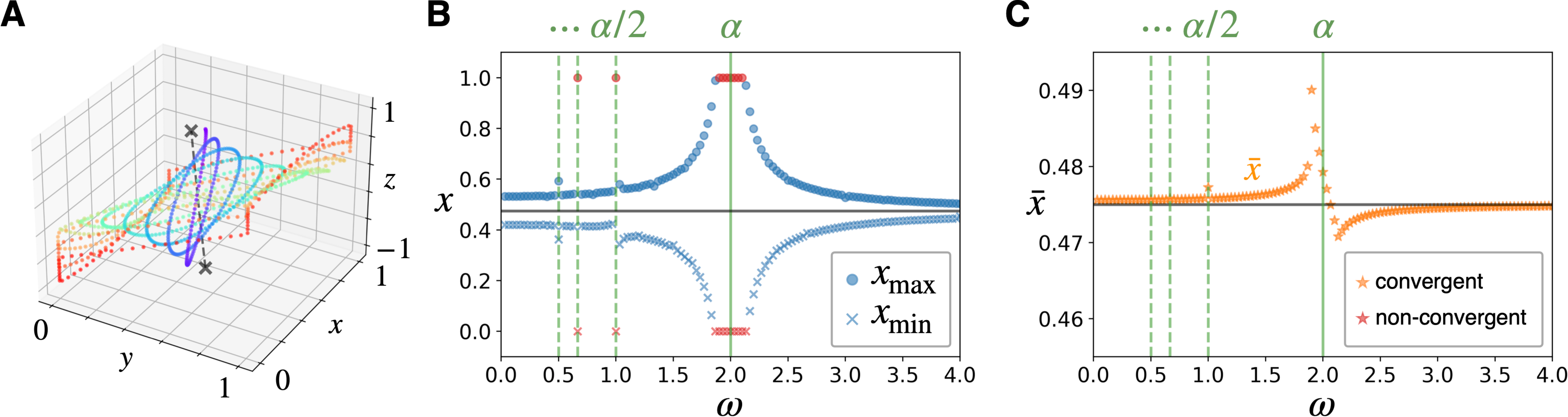}
    \caption{Learning dynamics under the boundary constraint of Eqs.~\eqref{x_FTRL_EU} and~\eqref{y_FTRL_EU}. All the parameters of the game and data are the same as Fig.~\ref{F02}. (A) The trajectory of learning dynamics in $\omega=\alpha$. The meanings of plots and axes are the same as Fig.~\ref{F01}-A. (B) The maximum value of $x(t)$ in overall time $0\le t\le T=10^{4}$. The meanings of the markers and axes are the same as Fig.~\ref{F02}-A. (C) The average value in the last time $T$. We regard the time-average as convergent when it moves less than $10^{-3}$ in the last $10^{2}$ time.The meanings of the markers and axes are the same as Fig.~\ref{F02}-B. The numerical method is the same as Fig.~\ref{F01} but the step size is $1/(4\times 10^{3})$.}
    \label{F04}
\end{figure*}

\subsection{Various Action Numbers}
\paragraph{$3\times 3$ matrix games:} We consider the case when each player can take three actions (see the upper panels in Fig.~\ref{F03}). Here, note that the periodic small perturbation of the payoff matrix is considered, which causes the slight vibration of the interior Nash equilibrium. As different from $2\times 2$ games, $3\times 3$ ones have multiple non-zero eigenvalues, $\alpha^{(1)}\ge\alpha^{(2)}>0$. From Panel B, we see that when $\alpha^{(j)}/\omega\in\mathbb{N}$ holds for either $j=1$ or $2$, the learning dynamics diverge from the Nash equilibrium, and otherwise, they cycle around the equilibrium. Panel C shows that when $\alpha^{(j)}/\omega\in\mathbb{N}$ holds, the time-average of the dynamics deviates from the Nash equilibrium, while otherwise, it converges to near the equilibrium.


\paragraph{$6\times 6$ matrix games:} Even if the game has a larger number of actions, similar results can be obtained. The lower panels of Fig.~\ref{F03} show an example of a $6\times 6$ matrix game with an interior equilibrium that fluctuates with time (see Panel A). This matrix game is equipped with five non-zero eigenvalues. Although the number of the eigenvalues is different, Panels B and C similarly demonstrate that the synchronization occurs only when $\alpha^{(j)}/\omega\in\mathbb{N}$ holds for either $j=1,2,3,4$ or $5$.

\paragraph{Intuition from experiments:} Intuitively, the above synchronization can occur in general two-player zero-sum games. As already explained, such games only have purely imaginary eigenvalues in general. Since the learning dynamics in general two-player zero-sum games are fixed in the Nash equilibrium, they are solved as the summation of multiple cycles around the equilibrium with frequencies $\alpha^{(j)}$. Thus, the oscillation of the equilibrium in a periodic game can resonate with these cycles. Furthermore, this synchronization is expected to occur in polymatrix games equipped with more than two players. These polymatrix games are known to have cycles in the learning dynamics. Indeed, we have observed the synchronization in three-player Matching Pennies, an example of polymatrix games (see Appendix~E.2).

\subsection{Boundary Constraint}
Next, we consider the boundary constraint given by the Euclidean regularizer in Eqs.~\eqref{x_FTRL_EU} and~\eqref{y_FTRL_EU}. The learning dynamics are different from Eq.~\eqref{cFTRL_EU} only on the boundary of the strategy spaces, i.e., $x, y\in\{0,1\}$ (see Panel A). Despite the existence of the boundary constraint, in $\alpha/\omega\in\mathbb{N}$ and its neighbor, one's strategy oscillates with a large amplitude (see Panel B). Elsewhere, the amplitude is small. However, since the amplitude is bounded by the boundary constraint, the time-average of one's strategy always converges, even in $\alpha/\omega\in\mathbb{N}$ (see Panel C).


\section{Conclusion}
This study examined learning in periodic games where the Nash equilibrium is time-varying. We discussed the dynamics of learning and their time-average convergence/non-convergence. Notably, this study focused on the synchronization of learning with the periodic change of games. We identified a phenomenon where the learning dynamics qualitatively change depending on whether or not the synchronization occurs. When the synchronization occurs, learning dynamics diverge from the Nash equilibrium, and their time average does not converge. Otherwise, the dynamics enter a complex cycle, but their time-average converges. We proved this phenomenon in a wide range of games, but with a limitation of $2\times 2$ matrices, the smooth periodicity of the game, and the absence of the boundary constraint of the strategy space. Our experiments demonstrated that this phenomenon universally occurs, regardless of these limitations.

One of the remaining problems is that the time-average converges elsewhere than the Nash equilibrium of the game's time-average. This problem is contrary to classical time-invariant games but consistent with~\cite{fiez2021online}. As a future work, it would be interesting to consider how to track the time-varying Nash equilibrium or converge to the Nash equilibrium of the time-average of periodic games. Another problem is to see and analyze the learning dynamics of other algorithms, such as optimistic~\cite{rakhlin2013optimization, syrgkanis2015fast, daskalakis2018training, daskalakis2019last, cai2022finite}, extra-gradient~\cite{korpelevich1976extragradient, mertikopoulos2019optimistic, lee2021fast, cai2022finite}, and negative momentum~\cite{polyak1964some, gidel2019negative, kovachki2021continuous, zhang2021suboptimality, hemmat2023lead}, where the strategies themselves converge to the Nash equilibrium, called last-iterate convergence. The behavior of such algorithms is already known under periodic games when the Nash equilibrium does not move~\cite{feng2023last, feng2024last}, but not when it moves. In addition, it would be interesting to analyze the synchronization in polymatrix games~\cite{bergman1998separable, cai2011minmax, bailey2019multi} more deeply. It is non-trivial how synchronization occurs when a polymatrix game can be divided into multiple games equipped with different frequencies and eigenvalues. This study finds a new phenomenon in learning in periodic games and gives insight into the phenomenon, providing a theoretical basis for these future problems.

\section*{Acknowledgments}
We thank Sosuke Ito for the fruitful discussions on linear dynamical systems. K. Ariu is supported by JSPS KAKENHI Grant No. 23K19986.


\appendix
\onecolumn

\setcounter{secnumdepth}{2} 

\begin{center}
{\Large\bf Appendix}
\end{center}

\section{Purely imaginary eigenvalues in FTRL with Euclidean regularizer} \label{App_purely}
This section is dedicated to proving that all the eigenvalues of $\bs{A}$ are zero or purely imaginary. Let $\lambda$ and $\bs{v}$ be a pair of an eigenvalue and an eigenvector, and then it holds
\begin{align}
    \lambda\bs{v}=\bs{A}\bs{v}=\begin{pmatrix}
        \bs{O}_{\sf X} & +\bs{P}_{\sf X}\bs{U} \\
        -\bs{P}_{\sf Y}\bs{U}^{\rm T} & \bs{O}_{\sf Y} \\
    \end{pmatrix}\begin{pmatrix}
        \bs{v}_{\sf X} \\
        \bs{v}_{\sf Y} \\
    \end{pmatrix}=\begin{pmatrix}
        +\bs{P}_{\sf X}\bs{U}\bs{v}_{\sf Y} \\
        -\bs{P}_{\sf Y}\bs{U}^{\rm T}\bs{v}_{\sf X} \\
    \end{pmatrix}.
\end{align}
Here, we used the description of $\bs{v}=(\bs{v}_{\sf X}, \bs{v}_{\sf Y})$. From this equation, we obtain
\begin{align}
    \lambda\bs{1}_{\sf X}^{\rm T}\bs{v}_{\sf X}&=\bs{1}_{\sf X}^{\rm T}\bs{P}_{\sf X}\bs{U}\bs{v}_{\sf Y}=\bs{1}_{\sf X}^{\rm T}\bs{U}\bs{v}_{\sf Y}-\frac{1}{m_{\sf X}}\underbrace{\bs{1}_{\sf X}^{\rm T}\bs{1}_{\sf X}}_{=m_{\sf X}}\bs{1}_{\sf X}^{\rm T}\bs{U}\bs{v}_{\sf Y}=0, \\
    -\lambda\bs{1}_{\sf Y}^{\rm T}\bs{v}_{\sf Y}&=\bs{1}_{\sf Y}^{\rm T}\bs{P}_{\sf Y}\bs{U}^{\rm T}\bs{v}_{\sf X}=\bs{1}_{\sf Y}^{\rm T}\bs{U}^{\rm T}\bs{v}_{\sf X}-\frac{1}{m_{\sf Y}}\underbrace{\bs{1}_{\sf Y}^{\rm T}\bs{1}_{\sf Y}}_{=m_{\sf Y}}\bs{1}_{\sf Y}^{\rm T}\bs{U}^{\rm T}\bs{v}_{\sf X}=0.
\end{align}
Thus, if $\lambda\neq 0$, then $\bs{1}_{\sf X}^{\rm T}\bs{v}_{\sf X}=\bs{1}_{\sf Y}^{\rm T}\bs{v}_{\sf Y}=0$ holds. Using this, we obtain
\begin{align}
    \bs{v}^{\rm H}\bs{A}\bs{v}&=\bs{v}_{\sf X}^{\rm H}\bs{P}_{\sf X}\bs{U}\bs{v}_{\sf Y}-\bs{v}_{\sf Y}^{\rm H}\bs{P}_{\sf Y}\bs{U}^{\rm T}\bs{v}_{\sf X} \\
    &=\bs{v}_{\sf X}^{\rm H}\bs{U}\bs{v}_{\sf Y}-\frac{1}{m_{\sf X}}\underbrace{\bs{v}_{\sf X}^{\rm H}\bs{1}_{\sf X}}_{=0}\bs{1}_{\sf X}^{\rm T}\bs{U}\bs{v}_{\sf Y}-\bs{v}_{\sf Y}^{\rm H}\bs{U}^{\rm T}\bs{v}_{\sf X}+\frac{1}{m_{\sf Y}}\underbrace{\bs{v}_{\sf Y}^{\rm H}\bs{1}_{\sf Y}}_{=0}\bs{1}_{\sf Y}^{\rm T}\bs{U}^{\rm H}\bs{v}_{\sf X} \\
    &=\bs{v}_{\sf X}^{\rm H}\bs{U}\bs{v}_{\sf Y}-\bs{v}_{\sf Y}^{\rm H}\bs{U}^{\rm T}\bs{v}_{\sf X}.
\end{align}
We also calculate its Hermitian conjugate as
\begin{align}
    (\bs{v}^{\rm H}\bs{A}\bs{v})^{\rm H}&=\bs{v}^{\rm H}\bs{A}^{\rm T}\bs{v} \\
    &=-\bs{v}_{\sf X}^{\rm H}\bs{U}\bs{P}_{\sf Y}\bs{v}_{\sf Y}+\bs{v}_{\sf Y}^{\rm H}\bs{U}^{\rm T}\bs{P}_{\sf X}\bs{v}_{\sf X} \\
    &=-\bs{v}_{\sf X}^{\rm H}\bs{U}\bs{v}_{\sf Y}+\frac{1}{m_{\sf Y}}\bs{v}_{\sf X}^{\rm H}\bs{U}\bs{1}_{\sf Y}\underbrace{\bs{1}_{\sf Y}^{\rm T}\bs{v}_{\sf Y}}_{=0}+\bs{v}_{\sf Y}^{\rm H}\bs{U}^{\rm T}\bs{v}_{\sf X}-\frac{1}{m_{\sf X}}\bs{v}_{\sf Y}^{\rm H}\bs{U}^{\rm T}\bs{1}_{\sf X}\underbrace{\bs{1}_{\sf X}^{\rm T}\bs{v}_{\sf X}}_{=0} \\
    &=-\bs{v}_{\sf X}^{\rm H}\bs{U}\bs{v}_{\sf Y}+\bs{v}_{\sf Y}^{\rm H}\bs{U}^{\rm T}\bs{v}_{\sf X} \\
    &=-\bs{v}^{\rm H}\bs{A}\bs{v}.
\end{align}
By the definition of eigenvalues $\lambda\|\bs{v}\|^{2}=\bs{v}^{\rm H}\bs{A}\bs{v}$ holds. Its Hermitian conjugate $\tilde{\lambda}\|\bs{v}\|^{2}=(\bs{v}^{\rm H}\bs{A}\bs{v})^{\rm H}$ holds, too, where $\tilde{\lambda}$ denotes the complex conjugate of $\lambda$. Thus, we derive
\begin{align}
    (\bs{v}^{\rm H}\bs{A}\bs{v})^{\rm H}=-\bs{v}^{\rm H}\bs{A}\bs{v}\ \Leftrightarrow\ \tilde{\lambda}=-\lambda,
\end{align}
meaning that all the eigenvalues of $\bs{A}$ are zero or purely imaginary.

\section{Derivation of equations}
\subsection{Derivation of gradient descent-ascent dynamics} \label{deriv_gda}
This section is dedicated to deriving the continuous form of the gradient descent-ascent, i.e., Eq.~\eqref{x_FTRL_EU}. First, in the argument of minimum, the extreme condition is satisfied as
\begin{align}
    \bs{x}^{\dagger}-\bs{x}=c\bs{1}_{\sf X}.
\end{align}
Here, from the constraint of $\bs{x}\in\Delta^{m_{\sf X}-1}$, $c$ is calculated as
\begin{align}
    &\bs{1}_{\sf X}^{\rm T}\bs{x}^{\dagger}-\underbrace{\bs{1}_{\sf X}^{\rm T}\bs{x}}_{=1}=c\underbrace{\bs{1}_{\sf X}^{\rm T}\bs{1}_{\sf X}}_{=m_{\sf X}}\quad \Leftrightarrow c=\frac{1}{m_{\sf X}}(\bs{1}_{\sf X}^{\rm T}\bs{x}^{\dagger}-1).
\end{align}
By substituting this equation, we obtain
\begin{align}
    \bs{x}&=\bs{x}^{\dagger}-\frac{\bs{1}_{\sf X}}{m_{\sf X}}(\bs{1}_{\sf X}^{\rm T}\bs{x}^{\dagger}-1)=\bs{P}_{\sf X}\bs{x}^{\dagger}+\frac{\bs{1}_{\sf X}}{m_{\sf X}}.
\end{align}
Finally, its time derivative is
\begin{align}
    \dot{\bs{x}}=\bs{P}_{\sf X}\dot{\bs{x}}^{\dagger}=\bs{P}_{\sf X}\bs{U}\bs{y}.
\end{align}
In a similar way (replace $\bs{P}_{\sf X}$ with $\bs{P}_{\sf Y}$ and $\bs{U}\bs{y}$ with $-\bs{U}^{\rm T}\bs{x}$), we calculate $\dot{\bs{y}}$ as
\begin{align}
    \dot{\bs{y}}=-\bs{P}_{\sf Y}\bs{U}^{\rm T}\bs{x}.
\end{align}
By combining these equations, we obtain Eq.~\eqref{cFTRL_EU}.

\subsection{Derivation of eigenvalues in two-action games} \label{deriv_eigen}
Let $\lambda$ denote the eigenvalues of Eq.~\eqref{matrix_A}, then we obtain
\begin{align}
    &\det \begin{pmatrix}
        \lambda\bs{I}_{\sf X} & -\bs{P}_{\sf X}\bs{U} \\
        \bs{P}_{\sf Y}\bs{U}^{\rm T} & \lambda\bs{I}_{\sf Y} \\
    \end{pmatrix}=0 \\
    &\Leftrightarrow\det(\lambda\bs{I}_{\sf Y})\det(\lambda\bs{I}_{\sf X}+\bs{P}_{\sf X}\bs{U}(\lambda\bs{I}_{\sf Y})^{-1}\bs{P}_{\sf Y}\bs{U}^{\rm T})=0 \\
    &\Leftrightarrow \lambda=0,\quad {\rm or}\ \det(\lambda^2\bs{I}_{\sf X}+\bs{P}_{\sf X}\bs{U}\bs{P}_{\sf Y}\bs{U}^{\rm T})=0.
    \label{eigenvalue}
\end{align}
Thus, we only have to consider the eigenvalues of $\bs{P}_{\sf X}\bs{U}\bs{P}_{\sf Y}\bs{U}^{\rm T}$, which is computed as
\begin{align}
    \bs{P}_{\sf X}\bs{U}\bs{P}_{\sf Y}\bs{U}^{\rm T}&=\frac{1}{4}\begin{pmatrix}
        +1 & -1 \\
        -1 & +1 \\
    \end{pmatrix}\begin{pmatrix}
        u_{11}(t) & u_{12}(t) \\
        u_{21}(t) & u_{22}(t) \\
    \end{pmatrix}\begin{pmatrix}
        +1 & -1 \\
        -1 & +1 \\
    \end{pmatrix}\begin{pmatrix}
        u_{11}(t) & u_{21}(t) \\
        u_{12}(t) & u_{22}(t) \\
    \end{pmatrix}
    \nonumber\\
    &=\frac{1}{4}(u_{11}(t)-u_{12}(t)-u_{21}(t)+u_{22}(t))\begin{pmatrix}
        +1 \\
        -1 \\
    \end{pmatrix}\otimes\begin{pmatrix}
        u_{11}(t)-u_{21}(t) \\
        u_{12}(t)-u_{22}(t) \\
    \end{pmatrix}.
\end{align}
Thus, by the tensor product representation of $\bs{P}_{\sf X}\bs{U}\bs{P}_{\sf Y}\bs{U}^{\rm T}$, the eigenvalues are trivially obtained as
\begin{align}
    &\det(\lambda^2 \bs{I}_{\sf X}+\bs{P}_{\sf X}\bs{U}\bs{P}_{\sf Y}\bs{U}^{\rm T})=0
    \nonumber\\
    &\Leftrightarrow \lambda^2=-\frac{1}{4}(u_{11}(t)-u_{12}(t)-u_{21}(t)+u_{22}(t))^2,\ 0
    \nonumber\\
    &\Leftrightarrow \lambda=\pm i\frac{1}{2}(u_{11}(t)-u_{12}(t)-u_{21}(t)+u_{22}(t)),\ \pm 0.
\end{align}
In addition, the eigenvalue of learning is described as
\begin{align}
    \alpha^{(1)}&=\frac{1}{2}(\bar{u}_{11}-\bar{u}_{12}-\bar{u}_{21}+\bar{u}_{22}), \\
    \alpha^{(2)}&=0,
\end{align}
which correspond to Eqs.~\eqref{ev1} and~\eqref{ev2}.

\subsection{Initial conditions of Eqs.~\eqref{x_sol_simplest1}-\eqref{y_sol_simplest2}} \label{deriv_simplest}
In Eqs.~\eqref{x_sol_simplest2} and~\eqref{y_sol_simplest2}, where $\alpha\neq\omega$ is assumed, $c_1$ and $c_2$ are uniquely determined by the initial state condition as
\begin{align}
    c_1=x(0)-\bar{x}^{*}-\frac{\alpha^2}{\alpha^2-\omega^2}\Delta x^{*},\quad c_2=y(0)-\bar{y}^{*}-\frac{\alpha^2}{\alpha^2-\omega^2}\Delta y^{*}.
\end{align}
Similarly, $c_1$ and $c_2$ are also uniquely determined in Eqs.~\eqref{x_sol_simplest1} and~\eqref{y_sol_simplest1} as
\begin{align}
    c_1=x(0)-\bar{x}^{*}-\frac{1}{2}\Delta x^{*},\quad c_2=y(0)-\bar{y}^{*}-\frac{1}{2}\Delta y^{*}.
\end{align}

\subsection{Derivation of Eqs.~\eqref{x_sol_general} and~\eqref{y_sol_general}} \label{deriv_general}
It is verified that Eqs.~\eqref{x_sol_general} and~\eqref{y_sol_general} are the solution of Eqs.~\eqref{dx_general} and~\eqref{dy_general} as follows.
\begin{align}
    \dot{x}(t)&=h(t)\underbrace{\sin(F(t)-F(t))}_{=0}+f(t)\int_{0}^{t}h(\tau)\cos(F(t)-F(\tau))\rmd\tau-g(t)\underbrace{\cos(F(t)-F(t))}_{=1}+f(t)\int_{0}^{t}g(\tau)\sin(F(t)-F(\tau))\rmd\tau
    \nonumber\\
    &\hspace{0.3cm}-f(t)c_1\sin F(t)+f(t)c_2\cos F(t)
    \nonumber\\
    &=+f(t)\underbrace{\left(\int_{0}^{t}g(\tau)\sin(F(t)-F(\tau))\rmd\tau+\int_{0}^{t}h(\tau)\cos(F(t)-F(\tau))\rmd\tau+c_2\cos F(t)-c_1\sin F(t)\right)}_{=y(t)}-g(t)
    \nonumber\\
    &=+f(t)y(t)-g(t), \\
    \dot{y}(t)&=g(t)\underbrace{\sin(F(t)-F(t))}_{=0}+f(t)\int_{0}^{t}g(\tau)\cos(F(t)-F(\tau))\rmd\tau+h(t)\underbrace{\cos(F(t)-F(t))}_{=1}-f(t)\int_{0}^{t}h(\tau)\sin(F(t)-F(\tau))\rmd\tau
    \nonumber\\
    &\hspace{0.3cm}-f(t)c_2\sin F(t)-f(t)c_1\cos F(t)
    \nonumber\\
    &=-f(t)\underbrace{\left(\int_{0}^{t}h(\tau)\sin(F(t)-F(\tau))\rmd\tau-\int_{0}^{t}g(\tau)\cos(F(t)-F(\tau))\rmd\tau+c_1\cos F(t)+c_2\sin F(t)\right)}_{=x(t)}+h(t)
    \nonumber\\
    &=-f(t)x(t)+h(t).
\end{align}

We also determine $c_1$ and $c_2$ by the initial state condition of $x(0)$ and $y(0)$ as
\begin{align}
    c_1=x(0)\cos F(0)-y(0)\sin F(0),\quad c_2=y(0)\cos F(0)+x(0)\sin F(0).
\end{align}

\section{Proofs}
\subsection{Proof of Theorem~\ref{Thm_div_simplest}} \label{Proof_div_simplest}
\begin{proof}
From Eq.~\eqref{x_sol_simplest1}, we derive $\bar{x}(T)$ as
\begin{align}
    \bar{x}(T)&=\frac{1}{T}\int_{0}^{T} \bar{x}^{*}+\frac{1}{2}\Delta x^{*}(\omega t\sin\omega t+\cos\omega t)-\frac{1}{2}\Delta y^{*}\omega t\cos\omega t+c_1\cos\omega t+c_2\sin\omega t\rmd t
    \nonumber\\
    &=\frac{1}{T}\left[\bar{x}^{*}t-\frac{1}{2}\Delta x^{*}\left(t\cos\omega t-\frac{2}{\omega}\sin\omega t\right)-\frac{1}{2}\Delta y^{*}\left(t\sin\omega t+\frac{1}{\omega}\cos\omega t\right)+\frac{c_1}{\omega}\sin\omega t-\frac{c_2}{\omega}\cos\omega t\right]_{0}^{T}
    \nonumber\\
    &=\bar{x}^{*}-\left(\frac{1}{2}\Delta x^{*}\cos\omega T+\frac{1}{2}\Delta y^{*}\sin\omega T\right)+O(T^{-1}).
    \label{x_ave_simplest1}
\end{align}
In the same way, we derive $\bar{y}(T)$ as
\begin{align}
    \bar{y}(T)&=\bar{y}^{*}-\left(\frac{1}{2}\Delta y^{*}\cos\omega T-\frac{1}{2}\Delta x^{*}\sin\omega T\right)+O(T^{-1}).
    \label{y_ave_simplest1}
\end{align}
Thus, both $\bar{x}(T)$ and $\bar{y}(T)$ no converge with $T\to\infty$ and continue to oscillate.
\end{proof}

\subsection{Proof of Theorem~\ref{Thm_conv_simplest}} \label{Proof_conv_simplest}
\begin{proof}
From Eq.~\eqref{x_sol_simplest2}, we derive $\bar{x}(T)$ as
\begin{align}
    \bar{x}(T)&=\frac{1}{T}\int_{0}^{T} \bar{x}^{*}+\frac{\alpha^2}{\alpha^2-\omega^2}\Delta x^{*}\cos\omega t+\frac{\alpha\omega}{\alpha^2-\omega^2}\Delta y^{*}\sin\omega t+c_1\cos\alpha t+c_2\sin\alpha t \rmd t
    \nonumber\\
    &=\frac{1}{T}\left[\bar{x}^{*}t+\frac{\alpha^2}{\omega(\alpha^2-\omega^2)}\Delta x^{*}\sin\omega t-\frac{\alpha}{\alpha^2-\omega^2}\Delta y^{*}\cos\omega t+\frac{c_1}{\alpha}\sin\alpha t-\frac{c_2}{\alpha}\cos\alpha t\right]_{0}^{T}
    \nonumber\\
    &=\bar{x}^{*}+O(T^{-1}).
\end{align}
Thus, we obtain $\lim_{T\to\infty}\bar{x}(T)=\bar{x}^{*}$. In the same way, we can prove $\lim_{T\to\infty}\bar{y}(T)=\bar{y}^{*}$.
\end{proof}

\subsection{Proof of Theorem~\ref{Thm_div_general}} \label{Proof_div_general}
\begin{proof}
We consider $\alpha/\omega=n\in\mathbb{N}$. Note that by definition, we need to calculate double integrals for $t$ and $\tau$ in order to derive the time-average $\bar{x}(T)$. The ranges of integrals, i.e., $t\in[0,T]$ and $\tau\in[0,t]$, are divided by the intervals of $2\pi/\omega$ (see Fig.~\ref{FA01}). We define
\begin{align}
    \mu(T):=\left\lfloor\frac{\omega T}{2\pi}\right\rfloor, \quad \nu(T):=T-\frac{2\pi}{\omega}\mu(T).
\end{align}
Here, $\mu(T)$ is the number of the intervals and is $O(T)$, while $\nu(T)$ is the remainder and is $O(1)$. Thus, $T=2\pi\mu(T)/\omega+\nu(T)$ holds.

Now, we describe the integral of the first term in Eq.~\eqref{x_sol_general} as
\begin{align}
    &\int_{0}^{T}\int_{0}^{t}h(\tau)\sin(F(t)-F(\tau))\rmd\tau\rmd t
    \nonumber\\
    &=\sum_{l=0}^{\mu(T)-1}\sum_{k=0}^{l-1}\int_{\frac{2\pi}{\omega}l}^{\frac{2\pi}{\omega}(l+1)}\int_{\frac{2\pi}{\omega}k}^{\frac{2\pi}{\omega}(k+1)}h(\tau)\sin(F(t)-F(\tau))\rmd\tau\rmd t+\int_{\frac{2\pi}{\omega}\mu(T)}^{\frac{2\pi}{\omega}\mu(T)+\nu(T)}\sum_{k=0}^{\mu(T)-1}\int_{\frac{2\pi}{\omega}k}^{\frac{2\pi}{\omega}(k+1)}h(\tau)\sin(F(t)-F(\tau))\rmd\tau\rmd t
    \nonumber\\
    &\hspace{0.3cm}+\sum_{l=0}^{\mu(T)-1}\int_{\frac{2\pi}{\omega}l}^{\frac{2\pi}{\omega}(l+1)}\int_{\frac{2\pi}{\omega}l}^{\frac{2\pi}{\omega}l+\nu(t)}h(\tau)\sin(F(t)-F(\tau))\rmd\tau\rmd t+\int_{\frac{2\pi}{\omega}\mu(T)}^{\frac{2\pi}{\omega}\mu(T)+\nu(T)}\int_{\frac{2\pi}{\omega}\mu(T)}^{\frac{2\pi}{\omega}\mu(T)+\nu(t)}h(\tau)\sin(F(t)-F(\tau))\rmd\tau\rmd t 
    \label{sync_division}\\
    &=\frac{\mu(T)(\mu(T)-1)}{2}\int_{0}^{\frac{2\pi}{\omega}}\int_{0}^{\frac{2\pi}{\omega}}h(\tau)\sin(F(t)-F(\tau))\rmd\tau\rmd t+\mu(T)\int_{0}^{\frac{2\pi}{\omega}}\int_{0}^{t}h(\tau)\sin(F(t)-F(\tau))\rmd\tau\rmd t
    \nonumber\\
    &\hspace{0.3cm}+\mu(T)\int_{0}^{\nu(T)}\int_{0}^{\frac{2\pi}{\omega}}h(\tau)\sin(F(t)-F(\tau))\rmd\tau\rmd t+\int_{0}^{\nu(T)}\int_{0}^{t}h(\tau)\sin(F(t)-F(\tau))\rmd\tau\rmd t.
\end{align}
Here, in the first equality, we divided the ranges of the integrals. In the second equality, we used the periodicity of $h(\tau)\sin(F(t)-F(\tau))$. In other words, $h(\tau)\sin(F(t)-F(\tau))$ is invariant about the transforms of $t\to t+\frac{2\pi}{\omega}l$ and $\tau\to \tau+\frac{2\pi}{\omega}k$ for all $k,l\in\mathbb{N}$ as
\begin{align}
    h(\tau+\frac{2\pi}{\omega}k)\sin(F(t+\frac{2\pi}{\omega}l)-F(\tau+\frac{2\pi}{\omega}k))&=h(\tau)\sin(F(t)-F(\tau)+2\pi\frac{\alpha}{\omega}(l-k))
    \nonumber\\
    &=h(\tau)\sin(F(t)-F(\tau)+2\pi n(l-k))
    \nonumber\\
    &=h(\tau)\sin(F(t)-F(\tau)).
\end{align}
Here, recall that $h(\cdot)$ and $\Delta F(\cdot)=F(\cdot)-\alpha t$ is periodic for frequency $\omega$.

Since $\mu(T)=O(T)$, the first term is $O(T^2)$ and dominant. Thus, it is proved that $\bar{x}(T)$ and $\bar{y}(T)$ diverge with the speeds of
\begin{align}
    \lim_{T\to\infty}\frac{1}{T}\bar{x}(T)&=\frac{1}{2}\left(\frac{\omega}{2\pi}\right)^2\int_{0}^{\frac{2\pi}{\omega}}\int_{0}^{\frac{2\pi}{\omega}}h(\tau)\sin(F(t)-F(\tau))-g(\tau)\cos(F(t)-F(\tau))\rmd\tau\rmd t,
    \label{x_div_term}\\
    \lim_{T\to\infty}\frac{1}{T}\bar{y}(T)&=\frac{1}{2}\left(\frac{\omega}{2\pi}\right)^2\int_{0}^{\frac{2\pi}{\omega}}\int_{0}^{\frac{2\pi}{\omega}}g(\tau)\sin(F(t)-F(\tau))+h(\tau)\cos(F(t)-F(\tau))\rmd\tau\rmd t.
    \label{y_div_term}
\end{align}

\paragraph{Interpretation of each term:}
Let us interpret the meaning of Eqs.~\eqref{sync_division} by Fig.~\ref{FA01}. The first terms of Eqs.~\eqref{sync_division} correspond to the red area of the figure, and the area is $O(T^2)$. By considering the average for $T$, its time-average is $O(T)$, thus called the ``divergence'' term. Next, the second terms correspond to the green area, and the area is $O(T)$. Thus, its time-average is $O(1)$. This area is also proportional to $\nu(T)$ and thus oscillates with frequency $\omega$, called the ``oscillation'' term. Next, the third terms correspond to the blue area, and the area is $O(T)$. Thus, its time-average is $O(1)$. Because this area converges to a value in the limit of $T\to\infty$, we call it the ``convergence'' term. Finally, the fourth terms correspond to the orange area, and the area is $O(1)$. Thus, its time-average is $O(1/T)$ and thus becomes negligible over time, called the ``negligible'' term.

\begin{figure}[h!]
    \centering
    \includegraphics[width=0.4\hsize]{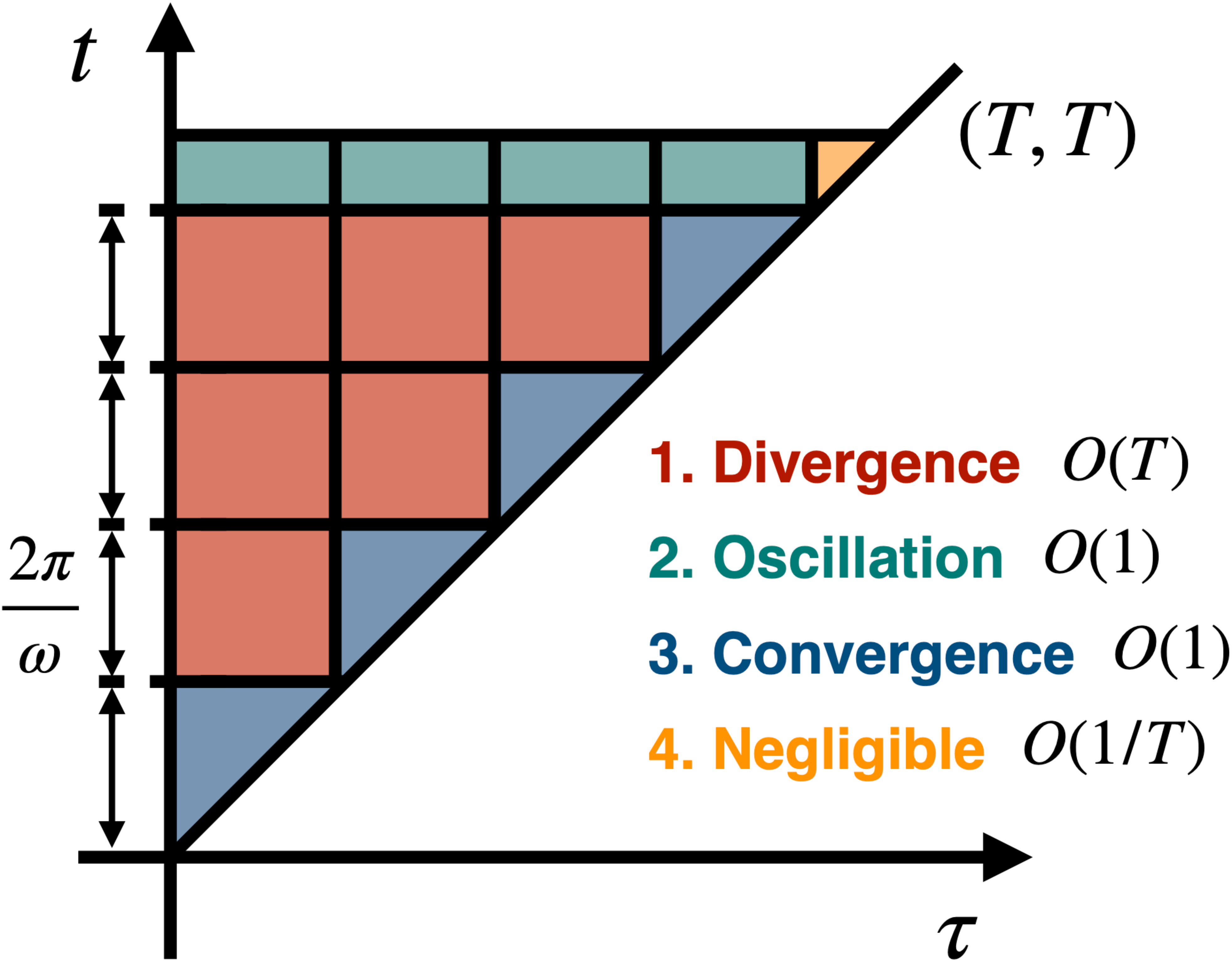}
    \caption{Schematics of the division of integral ranges. The x- and y-axes indicate the direction of $\tau$ and $t$, respectively. All the colored area (i.e., $0\le\tau\le t\le T$) shows the range of integral. Each square is of length $2\pi/\omega$. (1) The red area shows the divergence term, which is $O(T)$ in the time-average and thus diverges. (2) The green area shows the oscillation term, which is $O(1)$ in the time-average and oscillates over time. (3) The blue area shows the convergence term, which is $O(1)$ in the time-average. (4) Last, the orange area shows the negligible term, which is $O(1/T)$ in the time-average and thus disappears over time.}
    \label{FA01}
\end{figure}
\end{proof}

\subsection{Proof of Theorem~\ref{Thm_conv_general}} \label{Proof_conv_general}
\begin{proof}
We consider the case of $\alpha/\omega=r\in\mathbb{R}\setminus\mathbb{N}$. In this case, we divide the ranges of the integrals, i.e., $t\in[0,T]$ and $\tau\in[0,t]$, by the intervals of $2\pi/\omega$, again. We define the number of intervals $\mu(T)$ and the remainder $\nu(T)$ as
\begin{align}
    \mu(T):=\left\lfloor\frac{\omega T}{2\pi}\right\rfloor, \quad \nu(T):=T-\frac{2\pi}{\omega}\mu(T).
\end{align}

The third term of Eq.~\eqref{x_sol_general} is calculated as
\begin{align}
    &\int_{0}^{T}\int_{0}^{t}h(\tau)\sin(F(t)-F(\tau))\rmd\tau\rmd t
    \nonumber\\
    &=\underbrace{\sum_{l=0}^{\mu(T)-1}\sum_{k=0}^{l-1}\int_{\frac{2\pi}{\omega}l}^{\frac{2\pi}{\omega}(l+1)}\int_{\frac{2\pi}{\omega}k}^{\frac{2\pi}{\omega}(k+1)}h(\tau)\sin(F(t)-F(\tau))\rmd\tau\rmd t}_{=:({\rm A})}+\underbrace{\int_{\frac{2\pi}{\omega}\mu(T)}^{\frac{2\pi}{\omega}\mu(T)+\nu(T)}\sum_{k=0}^{\mu(T)-1}\int_{\frac{2\pi}{\omega}k}^{\frac{2\pi}{\omega}(k+1)}h(\tau)\sin(F(t)-F(\tau))\rmd\tau\rmd t}_{=:({\rm B})}
    \nonumber\\
    &\hspace{0.3cm}+\underbrace{\sum_{l=0}^{\mu(T)-1}\int_{\frac{2\pi}{\omega}l}^{\frac{2\pi}{\omega}(l+1)}\int_{\frac{2\pi}{\omega}l}^{t}h(\tau)\sin(F(t)-F(\tau))\rmd\tau\rmd t}_{=:({\rm C})}+\underbrace{\int_{\frac{2\pi}{\omega}\mu(T)}^{\frac{2\pi}{\omega}\mu(T)+\nu(T)}\int_{\frac{2\pi}{\omega}\mu(T)}^{t}h(\tau)\sin(F(t)-F(\tau))\rmd\tau\rmd t}_{=O(1)}
    \nonumber\\
    &=\frac{1}{2}\mu(T)\int_{0}^{\frac{2\pi}{\omega}}\int_{0}^{\frac{2\pi}{\omega}}h(\tau)\left(-\sin(F(t)-F(\tau))+\frac{\sin2\pi r}{1-\cos2\pi r}\cos(F(t)-F(\tau))\right)\rmd\tau\rmd t
    \nonumber\\
    &\hspace{0.3cm}+\mu(T)\int_{0}^{\frac{2\pi}{\omega}}\int_{0}^{t}h(\tau)\sin(F(t)-F(\tau))\rmd\tau\rmd t+O(1).
\end{align}

\paragraph{Divergence term:} The term $({\rm A})$ is calculated as
\begin{align}
    ({\rm A})&=\sum_{l=0}^{\mu(T)-1}\sum_{k=0}^{l-1}\int_{\frac{2\pi}{\omega}l}^{\frac{2\pi}{\omega}(l+1)}\int_{\frac{2\pi}{\omega}k}^{\frac{2\pi}{\omega}(k+1)}h(\tau)\sin(F(t)-F(\tau))\rmd\tau\rmd t
    \nonumber\\
    &=\sum_{l=0}^{\mu(T)-1}\sum_{k=0}^{l-1}\int_{0}^{\frac{2\pi}{\omega}}\int_{0}^{\frac{2\pi}{\omega}}h(\tau)\sin(F(t)-F(\tau)+2\pi r(l-k))\rmd\tau\rmd t
    \nonumber\\
    &=\sum_{l=0}^{\mu(T)-1}\sum_{k=0}^{l-1}\int_{0}^{\frac{2\pi}{\omega}}\int_{0}^{\frac{2\pi}{\omega}}h(\tau)(\sin(F(t)-F(\tau))\cos2\pi r(l-k)+\cos(F(t)-F(\tau))\sin2\pi r(l-k))\rmd\tau\rmd t
    \nonumber\\
    &=\frac{1}{2}\mu(T)\int_{0}^{\frac{2\pi}{\omega}}\int_{0}^{\frac{2\pi}{\omega}}h(\tau)\left(-\sin(F(t)-F(\tau))+\frac{\sin2\pi r}{1-\cos2\pi r}\cos(F(t)-F(\tau))\right)\rmd\tau\rmd t+O(1).
\end{align}
Here, we used
\begin{align}
    &\sum_{l=0}^{\mu(T)-1}\sum_{k=0}^{l-1}\cos2\pi r(l-k)+i\sum_{l=0}^{\mu(T)-1}\sum_{k=0}^{l-1}\sin2\pi r(l-k)=
    \sum_{l=0}^{\mu(T)-1}\sum_{k=0}^{l-1}e^{i2\pi r(l-k)}
    \nonumber\\
    &=\sum_{l=0}^{\mu(T)-1}\frac{e^{i2\pi rl}-1}{1-e^{-i2\pi r}}=\frac{1}{1-e^{-i2\pi r}}\left(\frac{1-e^{i2\pi r\mu(T)}}{1-e^{i2\pi r}}-\mu(T)\right)=\frac{-1}{1-e^{-i2\pi r}}\mu(T)+O(1)
    \nonumber\\
    &=\frac{-(1-\cos 2\pi r)+i\sin 2\pi r}{2(1-\cos 2\pi r)}\mu(T)+O(1)
    \nonumber\\
    &\Leftrightarrow \sum_{l=0}^{\mu(T)-1}\sum_{k=0}^{l-1}\cos2\pi r(l-k)=-\frac{1}{2}\mu(T)+O(1),\quad \sum_{l=0}^{\mu(T)-1}\sum_{k=0}^{l-1}\sin2\pi r(l-k)=\frac{\sin2\pi r}{2(1-\cos2\pi r)}\mu(T)+O(1)
\end{align}
Thus, in its time-average, the divergence term is $O(T)$ and is bounded.

\paragraph{Oscillation term:} The term $({\rm B})$ is also calculated as
\begin{align}
    ({\rm B})&=\int_{\frac{2\pi}{\omega}\mu(T)}^{\frac{2\pi}{\omega}\mu(T)+\nu(T)}\sum_{k=0}^{\mu(T)-1}\int_{\frac{2\pi}{\omega}k}^{\frac{2\pi}{\omega}(k+1)}h(\tau)\sin(F(t)-F(\tau))\rmd\tau\rmd t
    \nonumber\\
    &=\int_{\frac{2\pi}{\omega}\mu(T)}^{\frac{2\pi}{\omega}\mu(T)+\nu(T)}\sum_{k=0}^{\mu(T)-1}\int_{0}^{\frac{2\pi}{\omega}}h(\tau)\sin(F(t)-F(\tau)-2\pi rk)\rmd\tau\rmd t
    \nonumber\\
    &=O(1).
\end{align}
Here, we used
\begin{align}
    &\sum_{k=0}^{\mu(T)-1}\cos2\pi rk+i\sum_{k=0}^{\mu(T)-1}\sin2\pi rk=
    \sum_{k=0}^{\mu(T)-1}e^{i2\pi rk}=\frac{1-e^{i2\pi r\mu(T)}}{1-e^{i2\pi r}}=O(1)
    \nonumber\\
    &\Leftrightarrow\sum_{k=0}^{\mu(T)-1}\cos2\pi rk=O(1),\quad \sum_{k=0}^{\mu(T)-1}\sin2\pi rk=O(1).
\end{align}
Thus, the oscillation term can be ignored in the time-average.

\paragraph{Convergence term:} Next, the term $({\rm C})$ is calculated as
\begin{align}
    ({\rm C})&=\sum_{l=0}^{\mu(T)-1}\int_{\frac{2\pi}{\omega}l}^{\frac{2\pi}{\omega}(l+1)}\int_{\frac{2\pi}{\omega}l}^{\frac{2\pi}{\omega}l+\nu(t)}h(\tau)\sin(F(t)-F(\tau))\rmd\tau\rmd t
    \nonumber\\
    &=\sum_{l=0}^{\mu(T)-1}\int_{0}^{\frac{2\pi}{\omega}}\int_{0}^{t}h(\tau)\sin(F(t)-F(\tau))\rmd\tau\rmd t
    \nonumber\\
    &=\mu(T)\int_{0}^{\frac{2\pi}{\omega}}\int_{0}^{t}h(\tau)\sin(F(t)-F(\tau))\rmd\tau\rmd t.
\end{align}

Similarly, the integral of the fourth term of Eq.~\eqref{x_sol_general} is computed as
\begin{align}
    &-\int_{0}^{T}\int_{0}^{t}g(\tau)\cos(F(t)-F(\tau))\rmd\tau\rmd t
    \nonumber\\
    &=\frac{1}{2}\mu(T)\int_{0}^{\frac{2\pi}{\omega}}\int_{0}^{\frac{2\pi}{\omega}}g(\tau)\left(\frac{\sin2\pi r}{1-\cos2\pi r}\sin(F(t)-F(\tau))+\cos(F(t)-F(\tau))\right)\rmd\tau\rmd t
    \nonumber\\
    &\hspace{0.3cm}-\mu(T)\int_{0}^{\frac{2\pi}{\omega}}\int_{0}^{t}g(\tau)\cos(F(t)-F(\tau))\rmd\tau\rmd t+O(1).
\end{align}

We also consider the contribution of the initial state to the time-average. The third term of Eq.~\eqref{x_sol_general} is calculated as
\begin{align}
    \int_{0}^{T}\sin F(t)\rmd t&=\sum_{l=0}^{\mu(T)-1}\int_{\frac{2\pi}{\omega}l}^{\frac{2\pi}{\omega}(l+1)}\sin F(t)\rmd t+\underbrace{\int_{\frac{2\pi}{\omega}\mu(T)}^{\frac{2\pi}{\omega}\mu(T)+\nu(T)}\sin F(t)\rmd t}_{=O(1)}
    \nonumber\\
    &=\sum_{l=0}^{\mu(T)-1}\int_{0}^{\frac{2\pi}{\omega}}\sin F(t+\frac{2\pi}{\omega}l)\rmd t+O(1)
    \nonumber\\
    &=\sum_{l=0}^{\mu(T)-1}\int_{0}^{\frac{2\pi}{\omega}}\sin (F(t)+2\pi rl))\rmd t+O(1)
    \nonumber\\
    &=O(1),
\end{align}
Here, we used
\begin{align}
    &\sum_{l=0}^{\mu(T)-1}\cos2\pi rl+i\sum_{l=0}^{\mu(T)-1}\sin2\pi rl=\sum_{l=0}^{\mu(T)-1}e^{i2\pi rl}=\frac{1-e^{i2\pi\mu(T)}}{1-e^{i2\pi r}}=O(1)
    \nonumber\\
    &\Leftrightarrow \sum_{l=0}^{\mu(T)-1}\cos2\pi rl=O(1),\quad \sum_{l=0}^{\mu(T)-1}\sin2\pi rl=O(1).
\end{align}
Thus, the time-average does not depend on the initial state ($x(0)$ and $y(0)$).

We can derive the time-average for Eq.~\eqref{y_sol_general} similarly to the above. In conclusion, the limits of $\bar{x}(T)$ and $\bar{y}(T)$ converge to
\begin{align}
    \lim_{T\to\infty}\bar{x}(T)=\frac{\omega}{2\pi}&\left\{\int_{0}^{\frac{2\pi}{\omega}}\int_{0}^{t}h(\tau)\sin(F(t)-F(\tau))-g(\tau)\cos(F(t)-F(\tau))\rmd\tau\rmd t\right.
    \nonumber\\
    &\hspace{0.3cm}-\frac{1}{2}\int_{0}^{\frac{2\pi}{\omega}}\int_{0}^{\frac{2\pi}{\omega}}h(\tau)\sin(F(t)-F(\tau))-g(\tau)\cos(F(t)-F(\tau))\rmd\tau\rmd t
    \nonumber\\
    &\hspace{0.3cm}+\left.\frac{\sin2\pi r}{2(1-\cos2\pi r)}\int_{0}^{\frac{2\pi}{\omega}}\int_{0}^{\frac{2\pi}{\omega}}h(\tau)\sin(F(t)-F(\tau))+g(\tau)\cos(F(t)-F(\tau))\rmd\tau\rmd t\right\}, \\
    \lim_{T\to\infty}\bar{y}(T)=\frac{\omega}{2\pi}&\left\{\int_{0}^{\frac{2\pi}{\omega}}\int_{0}^{t}g(\tau)\sin(F(t)-F(\tau))-h(\tau)\cos(F(t)-F(\tau))\rmd\tau\rmd t\right.
    \nonumber\\
    &\hspace{0.3cm}-\frac{1}{2}\int_{0}^{\frac{2\pi}{\omega}}\int_{0}^{\frac{2\pi}{\omega}}g(\tau)\sin(F(t)-F(\tau))+h(\tau)\cos(F(t)-F(\tau))\rmd\tau\rmd t
    \nonumber\\
    &\hspace{0.3cm}+\left.\frac{\sin2\pi r}{2(1-\cos2\pi r)}\int_{0}^{\frac{2\pi}{\omega}}\int_{0}^{\frac{2\pi}{\omega}}g(\tau)\sin(F(t)-F(\tau))-h(\tau)\cos(F(t)-F(\tau))\rmd\tau\rmd t\right\}.
\end{align}
We have proved the time-average convergence.
\end{proof}

\section{Discussion of generic divergence in Thm.~\ref{Thm_div_general}} \label{generic_divergence}
This section discusses an exceptional case for the genetic divergence of the time-average in Thm.~\ref{Thm_div_general}. At a first glance, Thm.~\ref{Thm_div_general} looks inconsistent with Thm.~\ref{Thm_div_simplest}. Although the former shows that the time-average, i.e., $\bar{x}(T)$ and $\bar{y}(T)$, diverge with time, the latter shows that they cycle around the Nash equilibrium $(\bar{x}^{*},\bar{y}^{*})$. However, Thm.~\ref{Thm_div_simplest} is an exception where the divergence term, i.e., Eqs.~\eqref{x_div_term} and~\eqref{y_div_term}, is $0$. In general, we assume that two-action periodic games with a time-invariant eigenvalue. Then, it holds
\begin{align}
    f(t)=\alpha=n\omega \Leftrightarrow F(t)=n\omega t.
\end{align}
Thus, the divergence term is simply calculated as
\begin{align}
    &\int_{0}^{\frac{2\pi}{\omega}}\int_{0}^{\frac{2\pi}{\omega}}h(\tau)\sin(F(t)-F(\tau))-g(\tau)\cos(F(t)-F(\tau))\rmd\tau\rmd t
    \nonumber\\
    &=\int_{0}^{\frac{2\pi}{\omega}}\int_{0}^{\frac{2\pi}{\omega}}h(\tau)\sin n\omega(t-\tau)-g(\tau)\cos n\omega(t-\tau)\rmd\tau\rmd t
    \nonumber\\
    &=\int_{0}^{\frac{2\pi}{\omega}}h(\tau)\underbrace{\left[-\frac{1}{n\omega}\cos n\omega(t-\tau)\right]_{0}^{\frac{2\pi}{\omega}}}_{=0}-g(\tau)\underbrace{\left[\frac{1}{n\omega}\sin n\omega(t-\tau)\right]_{0}^{\frac{2\pi}{\omega}}}_{=0}\rmd\tau
    \nonumber\\
    &=0.
\end{align}
Here, in the second equality, we calculated the integral for $t$. Thus, the divergence term is proved to disappear. Instead of it, the oscillation and convergence terms become dominant, and Eqs.~\eqref{x_ave_simplest1} and~\eqref{y_ave_simplest1} are obtained.

\section{Other Experiments}
In this section, we provide experiments for other wider settings: 1) non-smooth wave shape and 2) polymatrix games. Even in these settings, we observe the synchronization phenomenon as follows.

\subsection{Non-smooth wave shape} \label{App_nonsmooth}
\begin{figure*}[h!]
    \centering
    \includegraphics[width=0.9\hsize]{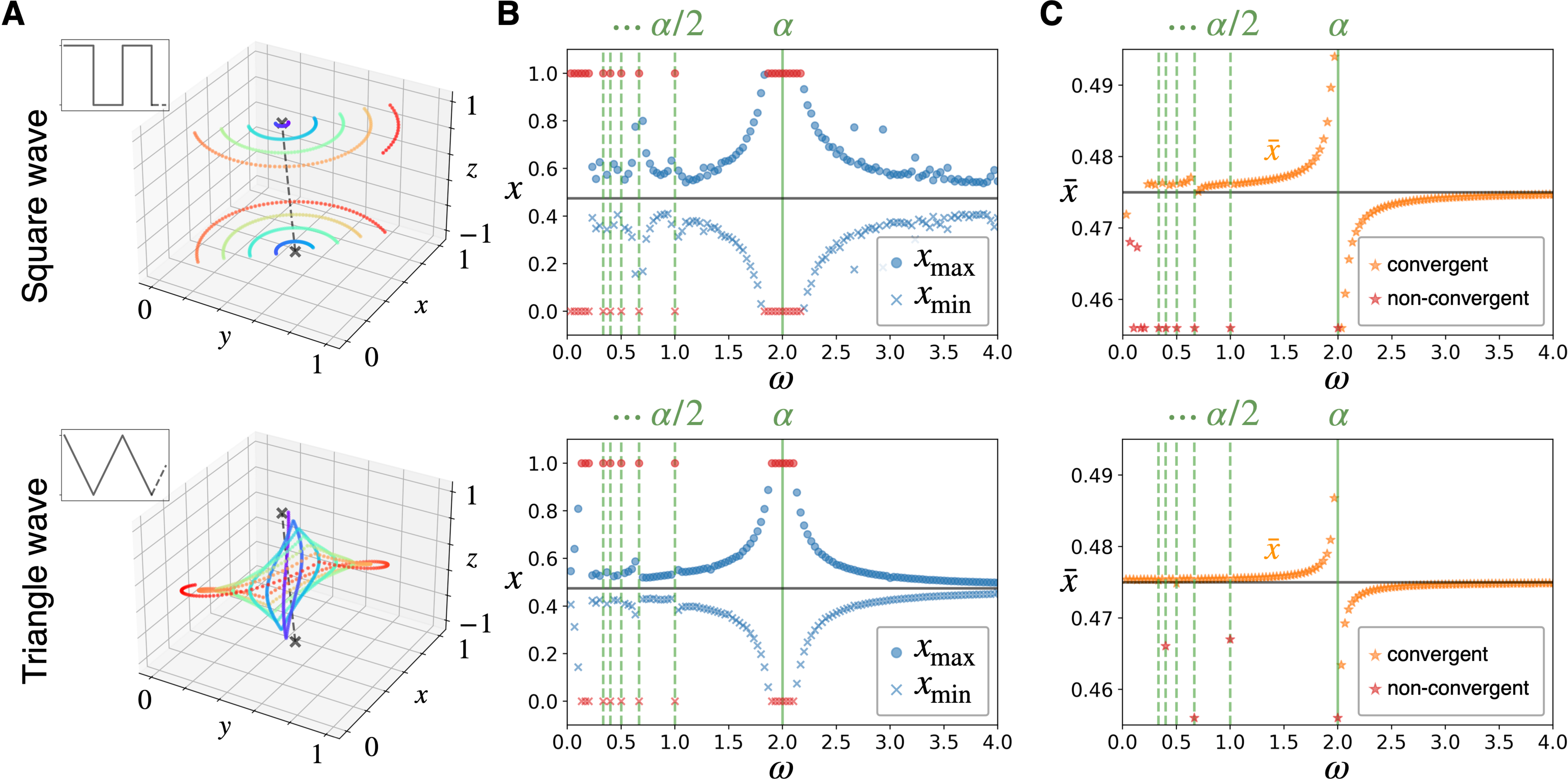}
    \caption{Learning dynamics for non-smooth game changes. The upper panels show square waves, while the lower panels show triangle waves. All the parameters of the game and data are the same as Fig.~\ref{F02}. (A) The trajectory of learning dynamics in $\omega=\alpha$. The meanings of plots and axes are the same as Fig.~\ref{F01}-A. (B) The maximum value of $x(t)$ for sufficiently large $t$. The meanings of the markers and axes are the same as Fig.~\ref{F02}-A. (C) The average value of $\bar{x}_1(T)$ for sufficiently large $T$. The meanings of the markers and axes are the same as Fig.~\ref{F02}-B.}
    \label{FA02}
\end{figure*}

First, Fig.~\ref{FA02} shows the case that games non-smoothly change with time. The upper panels show a square wave, which is a discontinuous function of $t$, while the lower ones show a triangle wave, which is a non-differentiable function (see Panels A for examples of typical trajectories). A similar result holds to the smooth periodic games (compare Panels B and C with Fig.~\ref{F02}). The only difference is that one's strategy and its time-average are more likely to diverge when $\alpha/\omega$ takes a large integer (see the left sides of Panels B and C). This is probably because the square and triangle waves have an infinite number of frequency components, i.e., $\omega, 2\omega, \cdots$. In conclusion, smoothness is not essential for our results but is used only to give the integral values.


\begin{figure*}[h!]
    \centering
    \includegraphics[width=0.9\hsize]{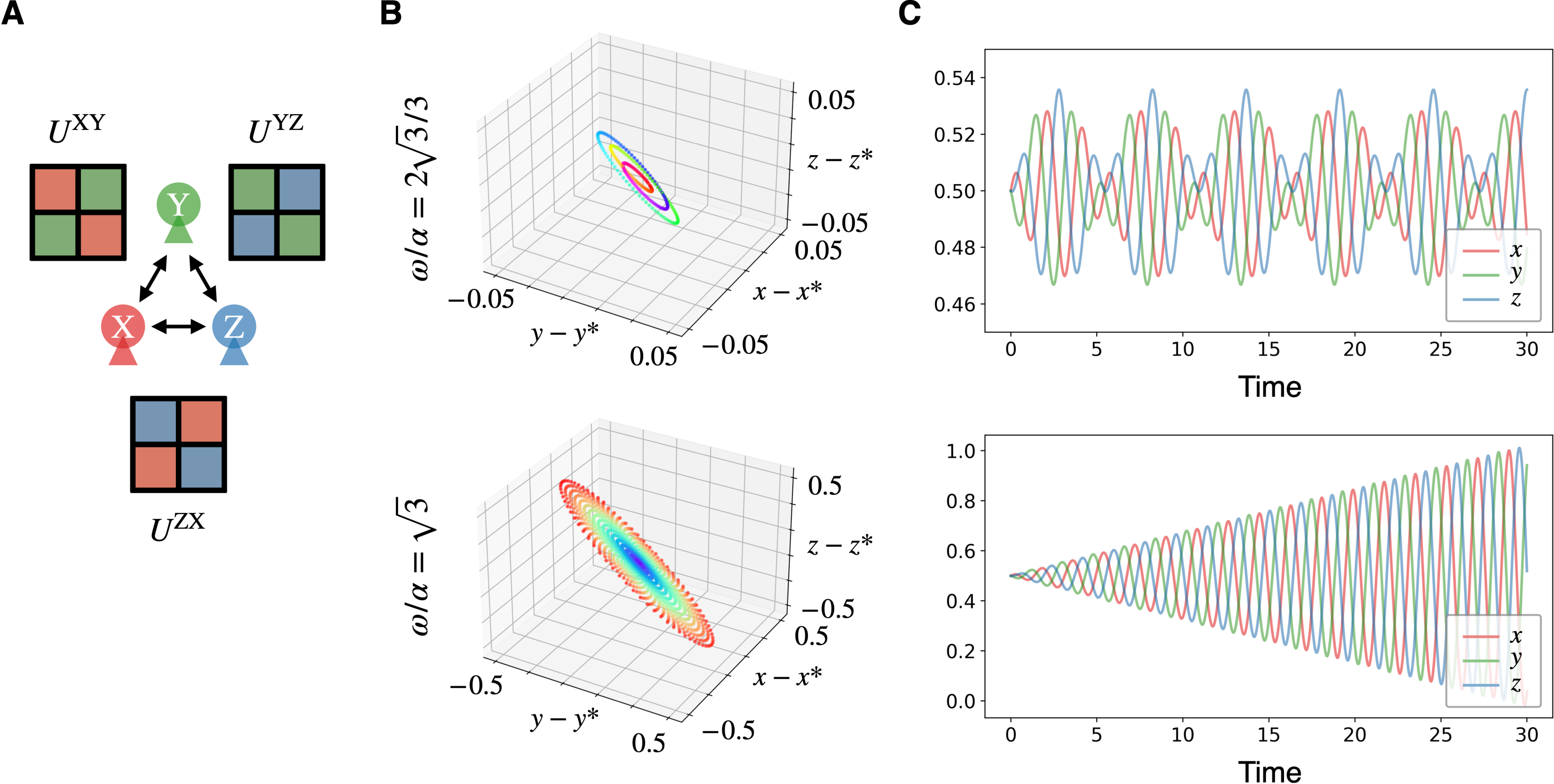}
    \caption{Learning dynamics in a game with more than two players. (A) Schematics of the Matching Pennies among the three players, X (red), Y (green), and Z (blue). The payoff matrix between X and Y is illustrated as $\bs{U}^{\rm XY}$, between Y and Z as $\bs{U}^{\rm YZ}$, and between Z and X as $\bs{U}^{\rm ZX}$. The row and column indicate the actions of the former and latter players, respectively. Each element of the payoff matrices is colored red, green, or blue, which means that the corresponding player (i.e., X, Y, and Z, respectively) obtains the unit score $1$. (B) The trajectory of the difference between their strategies ($x$, $y$, and $z$) and the Nash equilibrium ($x^{*}$, $y^{*}$, and $z^{*}$) is plotted for the two cases of $\omega/\alpha=2\sqrt{3}/3$ and $\sqrt{3}$. Here, note that because the game has continuous Nash equilibria, we measure the distance from the nearest Nash equilibrium. In the former case, the rainbow color shows the time of a single cycle. In the latter case, the rainbow color shows the passing of time from blue to red, meaning that the dynamics diverge from the equilibrium. (C) The time series of $x$ (red), $y$ (green), and $z$ (blue) for the two cases. The method (the fourth-order Runge-Kutta) and parameter (step size $=1/40$) are the same as those in Fig.~\ref{F02}.
    }
    \label{FA03}
\end{figure*}

\subsection{Polymatrix games} \label{App_polymatrix}
This paper considers two-player games in the setting, but this section deals with games with more than two players and observes the same synchronization phenomenon (see Fig.~\ref{FA03}). We extend the Matching Pennies used in Fig.~\ref{F01} into the case of three players, X, Y, and Z (visualized in Panel A). X plays the game of payoff matrix $\bs{U}^{\rm XY}(t)$ with Y, Y plays $\bs{U}^{\rm YZ}(t)$ with Z, and Z plays $\bs{U}^{\rm ZX}(t)$ with X. Here, all of these matrices $\bs{U}^{\rm XY}(t)$, $\bs{U}^{\rm YZ}(t)$, and $\bs{U}^{\rm ZX}(t)$ are given by eigenvalue invariant games (see Exm.~\ref{Exm_simplest}) as
\begin{align}
    \bar{\bs{U}}^{\rm XY}&=\bar{\bs{U}}^{\rm YZ}=\bar{\bs{U}}^{\rm ZX}=\begin{pmatrix}
        +1 & -1 \\
        -1 & +1 \\
    \end{pmatrix}, \\
    \Delta\bs{U}^{\rm XY}&=\frac{1}{10}\begin{pmatrix}
        +1 & 0 \\
        0 & -1 \\
    \end{pmatrix},\quad \Delta\bs{U}^{\rm YZ}=\frac{1}{20}\begin{pmatrix}
        +1 & 0 \\
        0 & -1 \\
    \end{pmatrix},\quad \Delta\bs{U}^{\rm ZX}=\frac{1}{30}\begin{pmatrix}
        +1 & 0 \\
        0 & -1 \\
    \end{pmatrix}.
\end{align}
Here, note that they play the Matching Pennies with each other in the time-average (see $\bar{\bs{U}}^{\rm XY}$, $\bar{\bs{U}}^{\rm YZ}$, and $\bar{\bs{U}}^{\rm ZX}$). The learning dynamics in this game are described by three variables $x, y, z\in[0,1]$ in the same method as Sec.~\ref{Subsec_analysis_eigenvalue}. We also compute $\alpha=2$ following Eq.~\eqref{ev1}. Now, Panel B shows the trajectory of the difference between their strategies ($x$, $y$, and $z$) and the Nash equilibrium ($x^{*}$, $y^{*}$, and $z^{*}$). We remark that this panel is the three-player version of Fig.~\ref{F01}-B. Interestingly, as different from the two-action case, the synchronization occurs when $\omega/\alpha=\sqrt{3}$, not $\omega/\alpha=1$. This is because the frequency of the cycling behavior becomes $\sqrt{3}$ times larger by the interaction among the three players. Panel C shows the time series of $x$, $y$, and $z$ in detail. We see that when the synchronization occurs ($\omega/\alpha=\sqrt{3}$), the amplitude of the learning dynamics grows in proportion to time.


\section{Computational environment}
The simulations presented in this paper were conducted using the following computational environment.
\begin{itemize}
\item Operating System: macOS Monterey (version 12.4)
\item Programming Language: Python 3.11.3
\item Processor: Apple M1 Pro (10 cores)
\item Memory: 32 GB
\end{itemize}

\end{document}